\documentclass[aps,twocolumn,floats,prd,nofootinbib,superscriptaddress,10pt]{revtex4-1}

\usepackage{graphicx,amsmath,amssymb}
\usepackage[hidelinks]{hyperref}
\hypersetup{
colorlinks=true,
linkcolor=red,
urlcolor=blue,
citecolor=blue}
\usepackage{tabularx}

\AtBeginDocument{
    \newwrite\bibnotes
    \def\bibnotesext{Notes.bib}
    \immediate\openout\bibnotes=\jobname\bibnotesext
    \immediate\write\bibnotes{@CONTROL{REVTEX41Control}}
    \immediate\write\bibnotes{@CONTROL{
    apsrev41Control,author="08",editor="1",pages="1",title="0",year="1"}}
     \if@filesw
     \immediate\write\@auxout{\string\citation{apsrev41Control}}
    \fi
}

\begin{document}

\title{Semi-analytical approach to Ly$\alpha$ multiple-scattering in 21-cm signal simulations}

\author{Jordan Flitter}
\email{E-mail: jordan.flitter@sns.it}
\affiliation{Scuola Normale Superiore di Pisa, Piazza dei Cavalieri 7, 56126 Pisa, Italy}

\author{Julian B. Mu\~{n}oz}
\affiliation{Department of Astronomy, The University of Texas at Austin, 2515 Speedway, Stop C1400, Austin, Texas 78712, USA}
\affiliation{Cosmic Frontier Center, The University of Texas at Austin, Austin, TX 78712, USA}
\affiliation{Texas Center for Cosmology \& Astroparticle Physics, Austin, TX 78712, USA}

\author{Andrei Mesinger}
\affiliation{Scuola Normale Superiore di Pisa, Piazza dei Cavalieri 7, 56126 Pisa, Italy}
\affiliation{Department of Physics and Astronomy {\it ``Ettore Majorana"}, University of Catania, Via Santa Sofia 64, 95123  Catania, Italy}
\affiliation{Centro Nazionale ``High Performance Computing, Big Data and Quantum Computing''}
             
\begin{abstract}

A crucial physical quantity in determining the 21-cm signal during cosmic dawn is the inhomogeneous background of Ly$\alpha$ photons originating from the first galaxies. As these photons travel through the intergalactic medium (IGM), their scattering cross-section is often approximated as a delta function at resonance due to computational cost.  That is, photons with emitted wavelengths between Ly$\alpha$ and Ly$\beta$ are assumed to travel in straight lines until they redshift into the Ly$\alpha$ resonance.  However, due to the damping wing in the Ly$\alpha$ cross-section, this approximation fails as the frequency of the photon approaches the resonant frequency, resulting in multiple scatterings events that could be separated by non-negligible distances. These multiple scattering events effectively modify the intrinsic Ly$\alpha$ emissivity from galaxies. Some previous works studied this effect of Ly$\alpha$ multiple scattering by running computationally heavy radiative-transfer simulations.  However, robustly interpreting the cosmic 21cm signal requires exploring a large parameter space of astrophysical uncertainties, motivating more computationally-efficient approaches. Here we incorporate Ly$\alpha$ multiple scatterings in the public, semi-numerical simulation {\tt 21cmFAST}. To do so, we employ Monte Carlo simulations to study the trajectories of Ly$\alpha$ photons on different scales. We find that the distance distributions of Ly$\alpha$ photons with respect to the absorption point can be modeled as analytical functions that are governed by a single parameter. Upon implementing the distance distributions in {\tt 21cmFAST}, we find that the multiple scattering effect is important (about 50\% difference in the 21-cm power spectrum) only at high redshifts before the spin temperature is fully coupled to the kinetic temperature. Furthermore, we find that Ly$\alpha$ multiple scattering does not enhance Ly$\alpha$ heating, and that the combined effect is negligible,
especially under realistic X-ray heating scenarios.

\end{abstract}

\maketitle


\section{Introduction}

The cosmological 21-cm signal contains information that could potentially deepen our understanding on the star formation history~\citep{Barkana:2000fd, Barkana:2004vb, Pritchard:2011xb, Park:2018ljd, Park:2019aul, Qin:2020xyh, Qin:2020pdx, Munoz:2021psm}, the epoch of reionization (EoR) ~\citep{Madau:1996cs, McQuinn:2005hk, Furlanetto:2006jb, Lidz:2008ry, Greig:2017jdj, Greig:2019tcg, Greig:2020suk, Gagnon-Hartman:2025oxd}, the nature of dark matter~\citep{Barkana:2018lgd, Kovetz:2018zan, Munoz:2018pzp, Munoz:2018jwq, Flitter:2022pzf, Cang:2023bnl, Flitter:2023mjj, Lazare:2024uvj}, and other cosmological quantities~\citep{Lidz:2013tra, Bernal:2017nec, Munoz:2019fkt, Cang:2021owu, Sarkar:2022mdz, Cruz:2023rmo, Plombat:2024kla, Adi:2024ebl, Libanore:2025ack}. Motivated by this potential, a worldwide effort is underway to detect the 21-cm signal. Some collaborations, like EDGES~\citep{Bowman:2007su}, SARAS~\citep{Patra:2012yw, Singh:2017syr}, BIGHORNS~\citep{2015PASA...32....4S}, REACH~\citep{deLeraAcedo:2022kiu}, PRIzM~\citep{2019JAI.....850004P}, MIST~\citep{Monsalve:2023lvo} and RHINO~\citep{Bull:2024doy} aim to measure the background evolution of the global (sky-averaged) signal. Others, like HERA~\citep{DeBoer:2016tnn}, LOFAR~\citep{2013A&A...556A...2V, 2021A&A...652A..37E}, and NenuFAR~\citep{7136773}, focus on measuring the 21-cm power spectrum. In the future, SKA~\citep{Braun:2015zta, Braun:2019gdo} is expected to be sensitive enough to map out the intergalactic medium (IGM) in 3D throughout the EoR. All the aforementioned experiments are poised to measure the signal at the cosmic dawn and reionization epochs, but there are as well several proposals to probe the signal at the preceding dark ages epoch with lunar-based instruments~\citep{2021PSJ.....2...44B, 2023arXiv230110345B, 2023ExA....56..741S, 2024AAS...24326401K}.

During cosmic dawn, the 21-cm signal is expected to have an absorption feature that arises due to strong coupling between the spin temperature of the IGM and the gas kinetic temperature via the Wouthuysen-Field effect~\citep{1952AJ.....57R..31W, 1958PIRE...46..240F, Hirata:2005mz}. This coupling becomes possible as Ly$\alpha$ radiation from the first stars interacts with the hydrogen atoms in the IGM. The Ly$\alpha$ coupling is thus a partially non-local effect, in the sense that the strength of the coupling at a given point in the IGM does not only depend on the luminosity of the closest sources. For simplicity, many 21-cm simulators in the literature, including {\tt 21cmFAST}~\citep{Mesinger:2010ne, Murray:2020trn}, {\tt SimFast21}~\citep{Santos:2009zk}, and {\tt Zeus21}~\citep{Munoz:2023kkg, Cruz:2024fsv} have made the simplifying assumption that the Ly$\alpha$ cross section is a delta function at resonance.  This effectively means that photons originating between Ly$\beta$ and Ly$\alpha$ travel in straight-lines from the source galaxy to the point at which they interact with the IGM.  In this straight-line (SL) approximation, there is negligible distance between scattering events because scattering only occurs at the resonant frequency of Ly$\alpha$.  However, the damping wing of the Ly$\alpha$ cross section results in some photons scattering {\it before} they redshift into resonance.  As a result, they can travel non-negligible distances between successive scattering events.  Accounting for such Ly$\alpha$ multiple scattering (MS) might be needed to accurately model the anisotropies of the cosmic 21cm signal.

Several works in the literature have previously studied the Ly$\alpha$ MS problem. Ref.~\cite{Loeb:1999er} found an analytical solution to the spatial and spectral distribution of nearly Ly$\alpha$ photons in an homogeneous, expanding IGM with zero temperature. Recently, Ref.~\cite{Smith:2025jsc} have managed to generalize the distribution of Ref.~\cite{Loeb:1999er} to account for an IGM of finite temperature. In the context of the 21-cm signal, Ref.~\cite{Reis:2021nqf} (hereafter RFB21) performed Monte Carlo (MC) simulations to incorporate the effect of Ly$\alpha$ MS in their 21-cm simulator, {\tt 21cmSPACE}~\citep{Visbal:2012aw, Pochinda:2023uom}. The MC code of RFB21, presented initially in Ref.~\cite{Reis:2020hrw} and later known in the literature as {\tt SPINTER}, was compared in Ref.~\cite{Semelin:2023lhz} to the {\tt LICORICE} simulation~\citep{Semelin:2007rk}, where it was concluded that peculiar velocities are an important aspect in the study of the effect that Ly$\alpha$ MS has on the 21-cm signal. Similar conclusions were also obtained by Ref.~\cite{Mittal:2023xih}, which incorporated in their simulations more advanced physical effects, such as anisotropic scattering and atom recoil. While the works of Refs.~\cite{Semelin:2023lhz, Mittal:2023xih} present the current state-of-the-art 21-cm simulations that account for Ly$\alpha$ MS, they were carried on radiative transfer simulations which are too computationally expensive for parameter inference with respect to future data from 21-cm observatories. 

In this paper, we take a similar approach to RFB21, in the sense that our analysis is based on performing first MC simulations to deduce the radial distributions of Ly$\alpha$ photons with respect to the absorption point, and then use these distributions in a semi-numerical 21-cm simulator, in our case, {\tt 21cmFAST}\footnote{\href{https://github.com/21cmfast/21cmFAST}{github.com/21cmfast/21cmFAST}}. However, unlike {\tt SPINTER}, our publicly available MC simulator, named speedy Ly$\alpha$ ray tracing algorithm ({\tt SP$\alpha$RTA}\footnote{\href{https://github.com/jordanflitter/SPaRTA}{github.com/jordanflitter/SPaRTA}}), contains many of the physical features that were studied in Ref.~\cite{Mittal:2023xih}, including peculiar velocities and finite IGM temperature. {\tt SP$\alpha$RTA} works very fast, between seconds to minutes, since it doesn't work with a grid, but instead employs analytical results from linear perturbation theory to emulate the effect of peculiar velocity on the scattered photons. Thus, {\tt SP$\alpha$RTA} was not designed to replicate precisely the results of advanced radiative transfer simulations on shorter time scales, but as a tool for approximately simulating the physics of Ly$\alpha$ MS. In addition, in contrast to RFB21 that have used a very complicated fitting function with six free parameters, we develop an analytical formalism that allows us to identify the important scales in the problem, as well as to find the connecting limit to the SL approximation. We use {\tt SP$\alpha$RTA} to find that the radial distance distributions can be modeled very well with analytical beta distributions that are controlled by a single parameter. By using our analytical formalism to feed appropriately the beta distributions into {\tt 21cmFAST}, we find that Ly$\alpha$ MS does matter when considering the 21-cm power spectrum, but only at high redshifts, when Ly$\alpha$ coupling is not too efficient. On top of that, contrary to RFB21, we find that the combined effect of Ly$\alpha$ MS and Ly$\alpha$ heating has a negligible effect on the 21-cm signal, even when considering X-ray luminosity far below the inferred value from HERA~\citep{HERA:2021noe} and the James Webb space telescope~\citep[JWST,][]{Breitman2026prep}.

The remaining parts of this paper are organized as follows. In Sec.~\ref{sec: Formalism} we describe our window function formalism for incorporating the effect of Ly$\alpha$ MS in the evaluation of the 21-cm signal. In Sec.~\ref{sec: SPaRTA} we describe how {\tt SP$\alpha$RTA} works. We then show in Sec.~\ref{sec: Analytical window function} the results of {\tt SP$\alpha$RTA} and how they can be used to model analytically the window functions associated with Ly$\alpha$ MS. These window functions are then used in Sec.~\ref{sec: The effect of Lya multiple scattering on the 21cm signal} to show how Ly$\alpha$ MS alters the fluctuations of the 21-cm signal. We conclude in Sec.~\ref{sec: Conclusions}. In addition, this paper contains several appendices. In Appendix \ref{sec: Derivation of the finite shell window function} we derive the finite shell window function, given the infinitely thin shell window function. In Appendix \ref{sec: Conditional Gaussian random variables} we outline the basic equations of conditional Gaussian random variables and in Appendix \ref{sec: Cosmological correlation functions} we show how the cosmological correlation functions are computed in linear perturbation theory. In Appendix \ref{sec: Beyond linear perturbation theory} we go beyond linear perturbation theory to validate our conclusions. Finally, in Appendix \ref{sec: Implementation of M-MS in 21cmFAST} we provide more details on our implementation of the MS window function in {\tt 21cmFAST}.

Throughout this work, we adopt for the cosmological parameters the Planck18 best-fit values \citep{Planck:2018vyg}. All physical scales in this paper correspond to comoving scales.

\section{Formalism}\label{sec: Formalism}

The quantity of most interest in 21-cm signal studies is the brightness temperature $T_{21}$, defined as
\begin{equation}\label{eq: 1}
T_{21}=\frac{T_\gamma-T_s}{1+z}\left(1-\mathrm{e}^{-\tau_{21}}\right),
\end{equation}
where $T_\gamma$ is the radiation temperature, often modeled as the cosmic microwave background (CMB) temperature, $T_s$ is the spin temperature of the IGM, and $\tau_{21}$ is the 21-cm optical depth. The spin temperature indicates the ratio between hydrogen atoms at the hyperfine triplet level and the singlet level. In thermal equilibrium, it reads
\begin{equation}\label{eq: 2}
T_s^{-1}=\frac{x_\mathrm{CMB}T_\gamma^{-1}+x_\mathrm{coll}T_k^{-1}+x_\alpha T_\alpha^{-1}}{x_\mathrm{CMB}+x_\mathrm{coll}+x_\alpha},
\end{equation}
where $T_k$ is the gas kinetic temperature, $T_\alpha\approx T_k$ is the Ly$\alpha$ color temperature, and $x_\mathrm{CMB}$, $x_\mathrm{coll}$ and $x_\alpha$ are the CMB~\citep{Venumadhav:2018uwn}, collisional~\citep{Furlanetto:2006jb}, and Ly$\alpha$~\citep{Hirata:2005mz} coupling coefficients, respectively. The latter coefficient, $x_\alpha$ is of most importance in this work. As the dark ages come to an end, $x_\mathrm{coll}$ is negligible while $x_\mathrm{CMB}\approx1$. Thus, the magnitude of $x_\alpha$ determines if $T_s$ follows more $T_\gamma$ or $T_k$; when $x_\alpha \gg 1$, $T_s\approx T_\alpha\approx T_k$.

There are several physical fields that determine $x_\alpha$ but the most critical one is the flux of the absorbed Ly$\alpha$ photons in the IGM, $J_\alpha$. In fact, we can write $x_\alpha\propto J_\alpha/J_0$, where $J_0\simeq5.6\times10^{-12}\left(1+z\right)\,\mathrm{cm^{-2}s^{-1}Hz^{-1}sr^{-1}}$~\citep{Mittal:2020kjs}, and the factor of proportionality is of order unity. Thus, approximately, when $J_\alpha\gg J_0$, $T_s\approx T_k$, regardless of the \emph{exact} value of $J_\alpha$, nor its spatial fluctuations. This condition is expected to hold at lower redshifts when the star formation rate density (SFRD) is larger. However, the fluctuations pattern of $J_\alpha$ do determine the fluctuations of $T_{21}$ at higher redshifts.

The evaluation of $J_\alpha$ requires taking into account contributions from three physical mechanisms. Firstly, the interaction of ionizing radiation with the IGM produces free electrons that can excite hydrogen atoms to the $n=2$ level, which consequently results in the production of Ly$\alpha$ photons as the excited hydrogen atoms return to ground state~\citep{Mesinger:2010ne}. This physical mechanism however yields the smallest contribution to $J_\alpha$. The other two mechanisms are continuum photons and injected photons~\citep{Chen:2003gc}. Continuum photons are photons with initial frequency between Ly$\alpha$ and Ly$\beta$ that redshift towards Ly$\alpha$ as the Universe expands. In contrast, injected photons are photons with initial frequency higher than Ly$\beta$. As the frequency of the injected photons redshifts towards Ly$n$ and they are absorbed by the IGM, Ly$\alpha$ photons can be produced upon atomic cascades. The combined contribution of continuum and injected photons to $J_\alpha$ can be expressed as
\begin{flalign}\label{eq: 3}
&\nonumber J_{\alpha}\left(\mathbf{x},z\right)=\frac{\left(1+z\right)^{2}}{4\pi}\sum_{n=2}^{n_{\mathrm{max}}}f_{\mathrm{recycle}}\left(n\right)&
\\&\hspace{10mm}\times\int_{z}^{z_{\mathrm{max}}\left(n\right)}dz'\frac{c}{H\left(z'\right)} I^\mathrm{app}_{*}\left(\nu'\right)\epsilon_{*}^\mathrm{app}\left(\mathbf{x},z'\right),&
\end{flalign}
where $z$ is the absorption redshift in which $J_\alpha$ is evaluated, $z'>z$ is a dummy integration variable that corresponds to the retarded time of emission, $H\left(z'\right)$ is the Hubble parameter at $z'$, $1+z_\mathrm{max}\left(n\right)=\left(1+z\right)\left[1-\left(n-1\right)^{-2}\right]/\left(1-n^{-2}\right)$ is the highest redshift for which a photon of initial frequency of Ly$n$ (denoted as $\nu_n$) will not be absorbed in the IGM before redshift $z$, $f_\mathrm{recycle}\left(n\right)$ denotes that probability that a Ly$n$ photon will cause the emission of a Ly$\alpha$ photon via atomic cascades~\citep{Hirata:2005mz, Pritchard:2005an}, we set $n_\mathrm{max}=23$ for numerical considerations~\citep{Mesinger:2010ne} and $c$ is the speed of light. In Eq.~\eqref{eq: 3}, the emissivity field $\epsilon_*\left(\mathbf x,z\right)$ is a quantity that describes how many photons are emitted per volume per unit of time at a point $\mathbf{x}$, and it is weighted by the spectral energy distribution (SED) $I_*\left(\nu\right)$. Integrating the product of both quantities along the comoving distance element $dz'c/H\left(z'\right)$ therefore returns a quantity of flux units.  

While $I_*\left(\nu\right)$ and $\epsilon_*\left(\mathbf x,z\right)$ are defined at the \emph{source} frame, these quantities need to be evaluated at the IGM's {target} frame. The photon frequency that the IGM ``sees'' is different than the source frequency due to redshift and Doppler shift, $\nu'=\left(1+z'\right)\left(1+z\right)\left(1-v_{\mathrm{rel}}^{||}/c\right)^{-1}\nu_{n}$, where $v_\mathrm{rel}^{||}$ is the parallel component of the relative velocity between the source and the absorbing atom. Thus, the \emph{apparent} SED is
\begin{flalign}\label{eq: 4}
&\nonumber I^\mathrm{app}_{*}\left(\nu'\right)=\int_{-\infty}^{\infty}dv_{\mathrm{rel}}^{||}I_{*}\left(\nu'\left(v_{\mathrm{rel}}^{||},\nu_{n}\right)\right]f_{v}\left(v_{\mathrm{rel}}^{||}\right)&
\\&\hspace{13mm}\approx I_{*}\left(\frac{1+z'}{1+z}\nu_{n}\right),&
\end{flalign}
where $f_v\left(v_\mathrm{rel}^{||}\right)$ is the distribution of $v_\mathrm{rel}^{||}$ and we approximated $f_v\left(v_\mathrm{rel}^{||}\right)\approx\delta^\mathrm{D}\left(v_\mathrm{rel}^{||}\right)$, with $\delta^\mathrm{D}$ the Dirac delta function. The last approximation effectively states that $v_{\mathrm{rel}}^{||}/c\ll 1$. As we shall see in Sec.~\ref{sec: Analytical window function}, this approximation is justified (especially for flat SEDs).

\subsection{Window function}\label{subsec: Window function}

Similarly to the apparent SED, the emissivity needs to be evaluated at the target frame as well. Due to the finite speed of light, it is not the local, instantaneous emissivity that contributes to the Ly$\alpha$ flux, but rather past and distant values of the emissivity field. The \emph{apparent} emissivity
of all contributing sources to the flux at point $\mathbf x$ at redshift $z$ is thus given by
\begin{flalign}\label{eq: 5}
&\nonumber\epsilon_*^{\mathrm{app}}\left(\mathbf{x},z'\right)=\int d^{3}x'f\left(\mathbf{x}';\mathbf{x},z,z'\right)\epsilon_*\left(\mathbf{x}-\mathbf{x}',z'\right)&
\\&\hspace{16mm}=f\left(\mathbf{x};z,z'\right)*\,\epsilon_*\left(\mathbf{x},z'\right),&
\end{flalign}
where * denotes convolution operation and $f\left(\mathbf{x}';\mathbf{x},z,z'\right)$ is the spatial distribution of photons that were emitted at redshift $z'$ from point $\mathbf{x}'$ and were later absorbed at redshift $z$ and point $\mathbf x$. Note that in the second equality of Eq.~\eqref{eq: 5} we assumed homogeneity in the distribution function, $f\left(\mathbf{x}';\mathbf{x},z,z'\right)=f\left(\mathbf{x}';z,z'\right)$, i.e. the spatial distribution does not depend on the absorption coordinate $\mathbf x$. At low redshifts, this assumption is in fact not accurate at all; after reionization kicks in, absorption points surrounded by ionized regions see different distributions than points surrounded by neutral regions. In addition, temperature fluctuations may change the apparent cross-section that the photons see during their trajectory to the absorption point. We will see however in Sec.~\ref{sec: The effect of Lya multiple scattering on the 21cm signal} that these details have little importance when the brightness temperature is the physical quantity to be considered, on the relevant scales for observation. Note also that in Eq.~\eqref{eq: 3} we have separated for simplicity any spectral effects from the apparent emissivity, implying that the spatial distributions are the same for all Ly$n$ photons that are absorbed in the IGM. This assumption was also taken by RFB21, and while it may not hold for all Ly$n$ photons in general, we adopted this assumption as well and decided to leave its relaxation for future work.

The convolution in Eq.~\eqref{eq: 5} can be efficiently computed by multiplying the two fields in Fourier space. To this end, we define the \emph{window function} to be the Fourier transform of the spatial distribution,
\begin{flalign}\label{eq: 6}
&W\left(\mathbf{k};z,z'\right)\equiv\mathrm{FT}\left[f\left(\mathbf{x};z,z'\right)\right]=\int d^{3}x\,\mathrm{e}^{-i\mathbf{k\cdot x}}f\left(\mathbf{x};z,z'\right).&
\end{flalign}
One can then easily compute the convolution in Eq.~\eqref{eq: 5} by multiplying the emissivity field in Fourier space with the window function, and then transform the product back to real space. It is important to note that the unfiltered emissivity field that is considered in Eq.~\eqref{eq: 5} is evaluated at the retarded redshift $z'$ rather the absorption redshift $z$.

To simplify the problem further, we shall assume that the spatial distribution is isotropic and purely radial, i.e. $f\left(\mathbf{x};z,z'\right)\propto f_{R_\mathrm{SL}}\left(r\,;z\right)$, where $r\equiv|\mathbf x-\mathbf x'|$ and $R_\mathrm{SL}$ is the comoving distance between $z$ and $z'$,
\begin{flalign}\label{eq: 7}
&\nonumber R_{\mathrm{SL}}\left(z,z'\right)\equiv\int_{z}^{z'}\frac{cdz''}{H\left(z''\right)}&
\\&\hspace{15mm}\approx\frac{2c}{H_{0}\Omega_{m}^{1/2}}\left[\left(1+z\right)^{-1/2}-\left(1+z'\right)^{-1/2}\right],&
\end{flalign}
where the second equality is an excellent approximation of the integral for a flat Universe during matter domination (as is the case for the redshift $5\lesssim z\lesssim 35$ of interest). For a purely radial distribution function $f_{R_\mathrm{SL}}\left(r\,;z\right)$, the window function becomes
\begin{eqnarray}\label{eq: 8}
\nonumber W\left(k R_\mathrm{SL};z\right)&=&\frac{\int_{0}^{{R_\mathrm{SL}}}drr^{2}f_{R_\mathrm{SL}}\left(r\,;z\right)\frac{\sin kr}{kr}}{\int_{0}^{{R_\mathrm{SL}}}drr^{2}f_{R_\mathrm{SL}}\left(r\,;z\right)}
\\&=&\frac{\int_{0}^{1}dyy^{2}f_1\left(y;z\right)\frac{\sin kR_\mathrm{SL}y}{kR_\mathrm{SL}y}}{\int_{0}^{1}dyy^{2}f_{1}\left(y;z\right)},
\end{eqnarray}
where in the second equality we switched integration variables and defined 
\begin{equation}\label{eq: 9}
y\equiv \frac{r}{R_\mathrm{SL}}.
\end{equation}
Note that Eq.~\eqref{eq: 8} implicitly assumes that the radial distribution vanishes for $r>R_\mathrm{SL}$ (or equivalently for $y>1$). This reflects the fact that, for a flat geometry, the longest comoving distance that a photon can travel between $z'$ and $z$ is $R_{\mathrm{SL}}\left(z,z'\right)$.

For numerical purposes, Eq.~\eqref{eq: 3} is evaluated by considering a finite amount of $z'$ values. Then, if the window function of Eq.~\eqref{eq: 8} is used to filter the emissivity field, numerical artifacts can appear when $J_\alpha\left(\mathbf x,z\right)$ is evaluated. For that reason, shells of finite width have to be considered in the integration of Eq.~\eqref{eq: 3} and the \emph{cumulative} radial distribution needs to be taken into account,
\begin{equation}\label{eq: 10}
f_{R_\mathrm{SL}}\left(r\,;z\right)\to\frac{1}{R_\mathrm{o}-R_\mathrm{i}}\int_{R_\mathrm{i}}^{R_\mathrm{o}}dr'\,f_{r'}\left(r\,;z\right),
\end{equation}
where $R_\mathrm{i}$ and $R_\mathrm{o}$ are the inner and outer radii of a given shell. Here, it is implicitly assumed that $f_{R_\mathrm{SL}}\left(r\,;z\right)$ is normalized such that $\int_0^{R_\mathrm{SL}}dr\,f_{R_\mathrm{SL}}\left(r\,;z\right)=1$. One can confirm that the cumulative window function approaches $f_{R_\mathrm{SL}}\left(r\,;z\right)$ in the infinitely thin shell limit where $R_\mathrm{o}\to R_\mathrm{i}=R_\mathrm{SL}$. In Appendix \ref{sec: Derivation of the finite shell window function} we show that the window function that corresponds to the cumulative radial distribution can be expressed as
\begin{equation}\label{eq: 11}
\tilde W\left(kR_\mathrm{o},kR_\mathrm{i};z\right)=\frac{R_{\mathrm{o}}^{3}\,M\left(kR_{\mathrm{o}};z\right)-R_{\mathrm{i}}^{3}\,M\left(kR_{\mathrm{i}};z\right)}{R_{\mathrm{o}}^{3}-R_{i}^{3}},
\end{equation}
where 
\begin{equation}\label{eq: 12}
M\left(kR;z\right)\equiv\frac{3}{k^3R^3}\int_0^{kR}dxx^{2}W\left(x\,;z\right).
\end{equation}

\subsection{Example: straight-line approximation}\label{Example: straight-line approximation}
To relate our window function formalism with the equations found in the literature, let us consider the concrete example of the \emph{SL approximation}. Here, given $z$ and $z'$, it is assumed that all photons that were absorbed at point $\mathbf x$ were emitted from sources of comoving distance $R_\mathrm{SL}$ from $\mathbf x$. Thus, the radial distribution that corresponds to the SL scenario is $f_{R_\mathrm{SL}}\left(r\,;z\right)=\delta^\mathrm{D}\left(r-R_\mathrm{SL}\right)$. Plugging this distribution in Eq.~\eqref{eq: 8} then yields $W\left(kR_\mathrm{SL}\right)=\sin kR_\mathrm{SL}/kR_\mathrm{SL}$. Then, plugging this window function in Eq.~\eqref{eq: 12} results in the SL window function,
\begin{equation}\label{eq: 13}
M_\mathrm{SL}\left(kR;z\right)=\frac{3\left(\sin kR-kR\cos kR\right)}{k^3R^3},
\end{equation}
which is the Fourier transform of a top-hat filter in real space. Notice that this window function is redshift-independent, this will not be the case when we consider the multiple-scattering scenario.

Up until now, Eqs.~\eqref{eq: 11} and \eqref{eq: 13} were implemented in {\tt 21cmFAST} and {\tt Zeus21} (see e.g.~\cite{Davies:2025wsa, Munoz:2023kkg}), reflecting the implicit assumption that photons in the simulation travel in straight-lines. In the following two sections, we explain how the MS window function is derived. Readers who are less interested in these details can jump ahead to Sec.~\ref{sec: The effect of Lya multiple scattering on the 21cm signal}, where we show the effect of the MS window function on the 21-cm signal.


\section{Speedy Ly$\alpha$ ray tracing algorithm}\label{sec: SPaRTA}

In this work, we incorporate the effect of Ly$\alpha$ MS in 21-cm simulations by generalizing the SL window function of Eq.~\eqref{eq: 13}. In order to find the MS window function, we use the newly developed tool {\tt {SP$\alpha$RTA}}. This MC code works by tracking the trajectories of many photons as they scatter off hydrogen atoms in the simulation. Since {\tt {SP$\alpha$RTA}} does not work with a grid, it is very efficient compared to other ray tracing and radiative transfer codes --- it takes seconds or minutes to perform a single MC simulation, depending on whether peculiar velocities are taken into account. {\tt {SP$\alpha$RTA}} resembles somewhat the {\tt SPINTER} code but we have included in our MC simulation four physical effects that were not taken into account in {\tt SPINTER} and were mostly inspired from the work of Ref.~\cite{Mittal:2023xih}. They are:
\begin{itemize}
\item \textbf{Peculiar velocity}: we account for random peculiar velocities in the IGM by using linear perturbation theory and conditional Gaussian random variables (see more information at appendices \ref{sec: Conditional Gaussian random variables} and \ref{sec: Cosmological correlation functions}). In our gridless approach, we draw the conditional peculiar random velocity vector given only the past sample of the peculiar velocity vector along the photon trajectory. While a more precise simulation would include information from more samples, or larger scales, we find this scheme accurate enough since the correlation function of the peculiar velocity is very high on small scales, and it decreases with larger scales.
\item \textbf{Finite temperature}: due to random thermal motions of the hydrogen atoms, the temperature of the IGM changes the effective cross-section in the gas rest frame from a simple Lorentzian shape to a Voigt profile~\citep{Dijkstra:2014xta, Dijkstra:2017lio}, 
\begin{flalign}\label{eq: 14}
&\nonumber\hspace{10mm}\sigma_\alpha\left(\nu_\mathrm{app};T_k\right)=\frac{3\lambda_\alpha^2a_T}{2\sqrt\pi}H\left(x_T;a_T\right)&
\\&\hspace{14mm}\simeq5.9\times10^{-14}\left(\frac{10^{4}\,\mathrm{K}}{T_k}\right)^{1/2}H\left(x_T;a_{T}\right)\,\mathrm{cm^{2}}&
\end{flalign}
where $\lambda_\alpha\equiv c/\nu_\alpha$ is the wavelength of the Ly$\alpha$ line center, $a_T\equiv A_\alpha/\left(4\pi\Delta\nu_\mathrm{D}\right)$, with $A_\alpha\simeq6.25\times10^8\,\mathrm{s}^{-1}$ the Einstein coefficient for spontaneous Ly$\alpha$ emission, $\Delta\nu_\mathrm{D}\equiv\left(2k_\mathrm{B}T_k/m_\mathrm{H}c^2\right)^{1/2}\nu_\alpha$, $k_\mathrm{B}$ is the Boltzmann constant and $m_\mathrm{H}$ is the hydrogen atom mass. The Voigt function is defined as
\begin{equation}\label{eq: 15}
H\left(x_T;a_T\right)\equiv\frac{a_T}{\pi}\int_{-\infty}^{\infty}dy\frac{\mathrm{e}^{-y^2}}{\left(y-x_T\right)^2+a_T^2},
\end{equation}
where $x_T\equiv\left(\nu_\mathrm{app}-\nu_\alpha\right)/\Delta\nu_\mathrm{D}$ and $\nu_\mathrm{app}$ is the apparent frequency of the photon, as seen by a hydrogen atom in the gas rest frame.

In {\tt {SP$\alpha$RTA}}, we have a free parameter $T_k$ that characterizes the temperature of the entire IGM in the simulation. Of course, a more realistic simulation would account for the fluctuations in the temperature of the IGM, and not just its global value. Yet, we will see in Sec.~\ref{subsec: Beta distributions} that this detail is less relevant for the scales of our interest.  
\item \textbf{Anisotropic scattering}: we also account for the anisotropic scattering of the photons in the gas rest frame; in this frame, photons are more likely to scatter forward or backward rather than sideways. This is manifest through the distribution~\citep{Dijkstra:2007jh, Mittal:2023xih}
\begin{equation}\label{eq: 16}
\hspace{7mm}f_\mu\left(\mu\right)=\begin{cases}
\left(11+3\mu^2\right)/24 & \text{if }\left|\nu_\mathrm{app}-\nu_\alpha\right|<0.2\Delta\nu_\mathrm{D} \\
3\left(1+\mu^2\right)/8 & \text{otherwise}
\end{cases},
\end{equation}
where $\mu$ is the cosine of the angle between the outgoing and incoming direction of the
photon.
\item \textbf{Atom recoil}: due to energy transfer between the scattered photon and the hydrogen atom, the frequency of the former is modified (see denominator of Eq.~\ref{eq: 21}).
\end{itemize}

{\tt {SP$\alpha$RTA}} works \emph{backwards} in time, i.e. given the absorption redshift $z_\mathrm{abs}$, the simulation begins at the line center when the photon has an apparent frequency $\nu_\mathrm{app}=\nu_\alpha$ and is absorbed by the IGM. This is done for two reasons. Firstly, the required distribution in Eq.~\eqref{eq: 5} is with respect to the absorption point, not with respect to the source of emission. Secondly, we can treat each intermediate point in the MC simulation as a potential source of emission, thereby reducing considerably the required amount of simulated photons for analyzing the radial distributions --- we work with only $N_\mathrm{photons}=10^3$, whereas typical MC simulations require $N_\mathrm{photons}$ higher by at least two orders of magnitude~\citep{2020A&A...635A.154M, 2022MNRAS.517.1767C, 2024MNRAS.532.3643Y, Smith:2014kna}. Another adjustable parameter in {\tt {SP$\alpha$RTA}} is the size of the virtual grid's cell, which for this work is set to be $L_\mathrm{cell}=200\,\mathrm{kpc}$. We have verified that our results are not sensitive to our choices for $N_\mathrm{photons}$ and $L_\mathrm{cell}$, indicating convergence.

Besides $z_\mathrm{abs}$, $N_\mathrm{photons}$, $L_\mathrm{cell}$, $T_k$ and the usual cosmological parameters, {\tt {SP$\alpha$RTA}} has in addition two more free parameters. The first parameter is a physical one and represents the global ionization fraction of the IGM, $x_\mathrm{HI}$. While ignoring the patchiness of the IGM during reionization is of course a crude approximation, we will see in Sec.~\ref{sec: The effect of Lya multiple scattering on the 21cm signal} that the exact fluctuations pattern of $J_\alpha$ is irrelevant in determining the brightness temperature. The other parameter in {\tt {SP$\alpha$RTA}} determines the redshift of the photon in the first step of the simulation. In this work, we use $\Delta z/\left(1+z_\mathrm{abs}\right)=2\times10^{-4}$. This value was chosen in order not to begin the simulation in the core of the Voigt profile (for $T_k=10^4\,\mathrm{K}$), where the slope of the cross-section changes very rapidly and the trapezoidal integration rule fails. But besides the numerical considerations behind this choice, there is also a more physical reason. One can confirm (e.g. by plugging into Eq.~\ref{eq: 7}) that this redshift difference corresponds to scales of order $500\,{\left(\frac{0.143}{\Omega_{m}h^2}\right)^{1/2}}\left(\frac{10}{1+z_{\mathrm{abs}}}\right)^{1/2}\,\mathrm{kpc}$, which is smaller than the scales that {\tt 21cmFAST} resolves.

\subsection{Algorithm description}\label{subsec: Algorithm description}
{\tt {SP$\alpha$RTA}} does not work with a grid. Instead, it tracks \emph{cosmological points} along the photon's trajectory. These points, denoted by $\mathcal P$, are characterized by four physical quantities:
\begin{enumerate}
\item The position vector of the photon $\mathbf x$, as measured in a comoving frame centered at the absorption point.
\item The peculiar velocity of the gas in the IGM at $\mathbf x$, $\mathbf v$, as measured in a \emph{rotated} comoving frame that one of its axes is aligned with the current photon's trajectory. This unusual choice of a frame is useful for the evaluation of velocity correlation functions (see more details at Appendix \ref{sec: Cosmological correlation functions}).
\item The redshift $z$ that corresponds to the cosmological time when the photon arrived to point $\mathbf x$.
\item The \emph{apparent} frequency of the photon at point $\mathbf x$, $\nu_\mathrm{app}$, as seen by a hydrogen atom in the gas rest frame (this is in contrast with the \emph{cosmological} frequency of the photon, as measured in a comoving frame).
\end{enumerate}
We denote a scattering event with an index $i$ (e.g. $\mathcal P_i$), while intermediate points between successive scattering events are denoted with a prime (e.g. $\mathcal P'$).

Note that {\tt {SP$\alpha$RTA}} does not track the over-density $\delta$ of the IGM. While in principle one could draw $\delta$ from a Gaussian distribution according to cosmological perturbation theory, a difficulty arises when the correlation between $\delta$ and the peculiar velocity field is considered (we elaborate more on that in Appendix \ref{subsec: Consistency check}). We therefore adopt the conclusions of Ref.~\cite{Mittal:2023xih} which have shown that inhomogeneities in the IGM density have little effect in determining the Ly$\alpha$ flux, and we thus assume that the photon travels in a mean-density IGM.

Below we outline how the algorithm of {\tt {SP$\alpha$RTA}} works. $\mathcal N\left(\mu,\sigma\right)$ denotes a Gaussian distribution of mean $\mu$ and root-mean-square (RMS) $\sigma$, while $\mathcal U\left(a,b\right)$ denotes a uniform distribution on the interval $[a,b]$. The following steps are performed for any of the $N_\mathrm{photons}$ photons, each of which has its own unique random seed.
\begin{enumerate}
\item \textbf{Setting the origin}. The simulation begins at the origin, $\mathbf x_\mathrm{abs}=\mathbf 0$, at redshift $z_\mathrm{abs}$ where the photon has an apparent frequency of $\nu_\mathrm{app}=\nu_\alpha$. A peculiar velocity vector $\mathbf v_\mathrm{abs}$ is drawn from $\mathcal N\left(0,\sigma_\mathrm{abs}\right)$ distribution, where $\sigma_\mathrm{abs}$ is computed from linear perturbation theory (see more details at Appendix \ref{sec: Cosmological correlation functions}).
\item \textbf{First step}. At $z_1=z_\mathrm{abs}+\Delta z$: 
\begin{enumerate}
\item The parallel component of the relative thermal velocity, $v_{\mathrm{rel},T}^{||}$, is drawn from $\mathcal N\left(0,v_\mathrm{th}\right)$, where $v_\mathrm{th}\equiv c\Delta\nu_\mathrm{D}/\nu_\alpha$.
\item The apparent frequency of the photon is determined from taking into account both redshift (or more precisely, blueshift, since the simulation works backwards in time) and Doppler shift.
\begin{equation}\label{eq: 17}
\nu_1=\frac{1+z_1}{1+z_\mathrm{abs}}\left(1-\frac{v_{\mathrm{rel},T}^{||}}{c}\right)^{-1}\nu_\mathrm{app}
\end{equation}
\item Given $\nu_1$, each component of the position vector of the photon $\mathbf x_1$ is drawn from $\mathcal N\left(0,\sigma_\mathbf{x}\right)$ where\footnote{We note that Eq.~\eqref{eq: 18} was derived analytically by Ref.~\cite{Loeb:1999er} while assuming zero temperature. Recently, Ref.~\cite{Smith:2025jsc} managed to generalize this result while considering a finite temperature. Yet, because it is not clear how to numerically draw from their distribution, and since their distribution is not significantly different for the initial frequency shift considered in this work, we chose to work with the distribution of Ref.~\cite{Loeb:1999er}.} \citep{Loeb:1999er}
\begin{equation}\label{eq: 18}
\hspace{15mm}\sigma_\mathbf{x}\equiv\left(\frac{2}{9}\right)^{1/2}\left(\frac{\nu_1-\nu_\alpha}{\Delta\nu_*\left(z_\mathrm{abs}\right)}\right)^{3/2}R_*\left(z_\mathrm{abs}\right),
\end{equation}
where $\Delta\nu_*\left(z_\mathrm{abs}\right)$ and $R_*\left(z_\mathrm{abs}\right)$ are given by Eqs.~\eqref{eq: 23}-\eqref{eq: 24}.
\item Given the initial distance of the photon with respect to the origin, $r_\mathrm{ini}$, the peculiar velocity $\mathbf v_1$ is drawn from $\mathcal N\left(\mu_\mathrm{cond},\sigma_\mathrm{cond}\right)$, where $\mu_\mathrm{cond}$ and $\sigma_\mathrm{cond}$ depend on $\mathbf v_\mathrm{abs}$ through conditional Gaussian random variables theory (see more details at Appendix \ref{sec: Conditional Gaussian random variables}). Given $\mathbf v_1$, the parallel component of the relative peculiar velocity $v_\mathrm{rel}^{||}=v_1^{||}-v_\mathrm{abs}^{||}$ is computed. This is then used to give a small Doppler correction to $\nu_1$ (this Doppler correction is smaller than $\nu_1-\nu_\alpha$, as $\mathbf{v}_1$ and $\mathbf{v}_\mathrm{abs}$ are very correlated).
\end{enumerate}
\item \textbf{Initialization (first scattering point)}. $\mathcal P_i$ is initialized to be $\mathcal P_1$.
\item \textbf{After each scattering event $i$}:
\begin{enumerate}
\item \textbf{Initialization (optical depth integral)}. $\mathcal P'$ is initialized to be $\mathcal P_i$. In addition, we draw $\mathrm{e}^{-\tau_\mathrm{rnd}}$ from $\mathcal U\left(0,1\right)$ and initialize $\tau_i=0$.
\item \textbf{Computing the relative velocity}. Given $z'$ and $L_\mathrm{cell}$, the next redshift $z''$ is found by inverting Eq.~\eqref{eq: 7}. Given $z''$ and $\mathbf v'$, $\mathbf v''$ is found via conditional Gaussian random variables theory (see more details at Appendices \ref{sec: Conditional Gaussian random variables}-\ref{sec: Cosmological correlation functions}). The relative velocity between the two intermediate points is computed, $\mathbf v_\mathrm{rel}=\mathbf v''-\mathbf v'$.
\item \textbf{Computing the apparent frequency}. The apparent frequency of the photon (as seen in the gas rest frame) is updated via blueshift and Doppler shift,
\begin{equation}\label{eq: 19}
\nu''=\frac{1+z''}{1+z'}\left(1-\frac{v_\mathrm{rel}^{||}}{c}\right)^{-1}\nu'.
\end{equation}
\item \textbf{Computing the cross-section}. Given the apparent frequency $\nu''$, the Voigt cross-section $\sigma_\alpha\left(\nu'';T_k\right)$ is computed according to Eq.~\eqref{eq: 14}.
\item \textbf{Updating the optical depth}. The background hydrogen density is computed, $n_\mathrm{HI}\left(z''\right)=x_\mathrm{HI}n_\mathrm{H0}\left(1+z''\right)^3$, where $n_\mathrm{H0}=\left(1-Y_\mathrm{He}\right)\rho_\mathrm{crit}\Omega_b/m_\mathrm{H}$ is the comoving hydrogen number-density, $Y_\mathrm{He}$ is the helium mass fraction and $\rho_\mathrm{crit}$ is the cosmological critical density. Then, the optical depth $\tau_i$ is increased\footnote{In practice, we compute $\Delta\tau_i$ via the trapezoidal integration rule (for brevity, the text shows $\Delta\tau_i$ as computed via the right rectangular rule).} by $\Delta\tau_i=n_\mathrm{HI}\left(z''\right)\sigma_\alpha\left(\nu'',T_k\right)L_\mathrm{cell}/\left(1+z''\right)$. If $\tau_i<\tau_\mathrm{rnd}$ and $\nu''<\nu_\beta$, the code returns to step 4(b) while updating $\mathcal P''\to \mathcal P'$. Otherwise, the code proceeds to step 5.
\end{enumerate}
\item \textbf{Next scattering event}. The next scattering event is identified when $\tau_\mathrm{rnd}\geq\tau_i$, where the code sets $z_{i+1}=z'$. The displacement in the photon position $\Delta\mathbf x_i=\mathbf x_{i+1}-\mathbf x_{i}$ is computed in spherical coordinates given the comoving distance $R_\mathrm{SL}\left(z_i,z_{i+1}\right)$ (Eq.~\ref{eq: 7}) and the random scatter angles, $\mu_\mathrm{rnd}$ and $\phi_\mathrm{rnd}$. $\mu_\mathrm{rnd}$ is drawn from the distribution of Eq.~\eqref{eq: 16} while $\phi_\mathrm{rnd}$ is drawn from $\mathcal U\left(0,2\pi\right)$.
\begin{enumerate}
\item \textbf{Outgoing frequency modification}. In the gas rest frame, the frequency of the incoming photon in the $i+1$ scattering event is $\nu_{i+1}^\mathrm{in}=\nu_{i+1}$. The outgoing frequency $\nu_{i+1}^\mathrm{out}$ is modified by two effects. Firstly, assuming a coherent scattering in the hydrogen atom rest frame, the outgoing frequency is Doppler shifted due to changing frames from the atom rest frame to the gas rest frame. This requires drawing the parallel and perpendicular components of the thermal velocity of the hydrogen atom, $v_T^{||}$ and $v_T^{\perp}$. While $v_T^\perp$ is drawn from $\mathcal N\left(0,v_\mathrm{th}/\sqrt{2}\right)$, the distribution of $v_T^{||}$ is more complicated since the photon has a preference to scatter from atoms with $v_T^{||}$ that would reduce the apparent frequency shift of the photon with respect to the line center. Given the dimensionless frequency shift $x_T$, the resulting distribution for $v_T^{||}$ can be deduced from Bayes theorem; it is the product of the unconditional probability to draw a random $v_T^{||}$ (a Gaussian) and the conditional probability that a scattering event occurred given $v_T^{||}$, which is given by a biased Lorentzian cross-section. Requiring a normalized conditional probability for $v_T^{||}$ given $x_T$ then yields the following distribution\footnote{In order to draw from the distribution of Eq.~\eqref{eq: 20}, we implement the rejection method of Ref.~\cite{Zheng:2002qc} but with the comparison function of the {\tt RASCAS} code \citep{2020A&A...635A.154M}.}~\citep{Semelin:2007rk, Dijkstra:2017lio, 2020A&A...635A.154M, Mittal:2023xih},
\begin{flalign}\label{eq: 20}
&\nonumber \hspace{17mm} f_u\left(u_T\,|\,x_T\right)=&
\\& \hspace{25mm}\frac{1}{H\left(x_T;a_T\right)}\frac{a_T}{\pi}\frac{\mathrm{e}^{-u_T^{2}}}{\left(u_T-x_T\right)^{2}+a_T^{2}}.&
\end{flalign}
where $u_T\equiv v_T^{||}/v_\mathrm{th}$. The second effect that modifies the outgoing frequency is energy transfer from the scattered photon to the atom due to recoil of the latter. Taking into account both effects, the outgoing frequency of the photon is~\citep{Mittal:2023xih}
\begin{flalign}\label{eq: 21}
&\nonumber \hspace{17mm}\nu_{i+1}^\mathrm{out}=&
\\&\hspace{17mm}\frac{1+\left(\mu_\mathrm{rnd}-1\right)v_T^{||}/c+\left(1-\mu_\mathrm{rnd}^2\right)^{1/2}v_T^\perp/c}{1+\left(1-\mu_\mathrm{rnd}\right)h_\mathrm{P}\nu_{i+1}^\mathrm{in}/\left(m_\mathrm{H}c^2\right)}\nu_{i+1}^\mathrm{in}.&
\end{flalign}
\end{enumerate}
\item \textbf{Termination}. If $\nu_{i+1}<\nu_\beta$ the code returns to step 4 with $\mathcal P_{i+1}\to \mathcal P_i$. Otherwise, the simulation is finished for this photon.
\end{enumerate}

\subsection{Case study: five photon trajectories}\label{subsec:}

\begin{figure*}
\includegraphics[width=\textwidth]{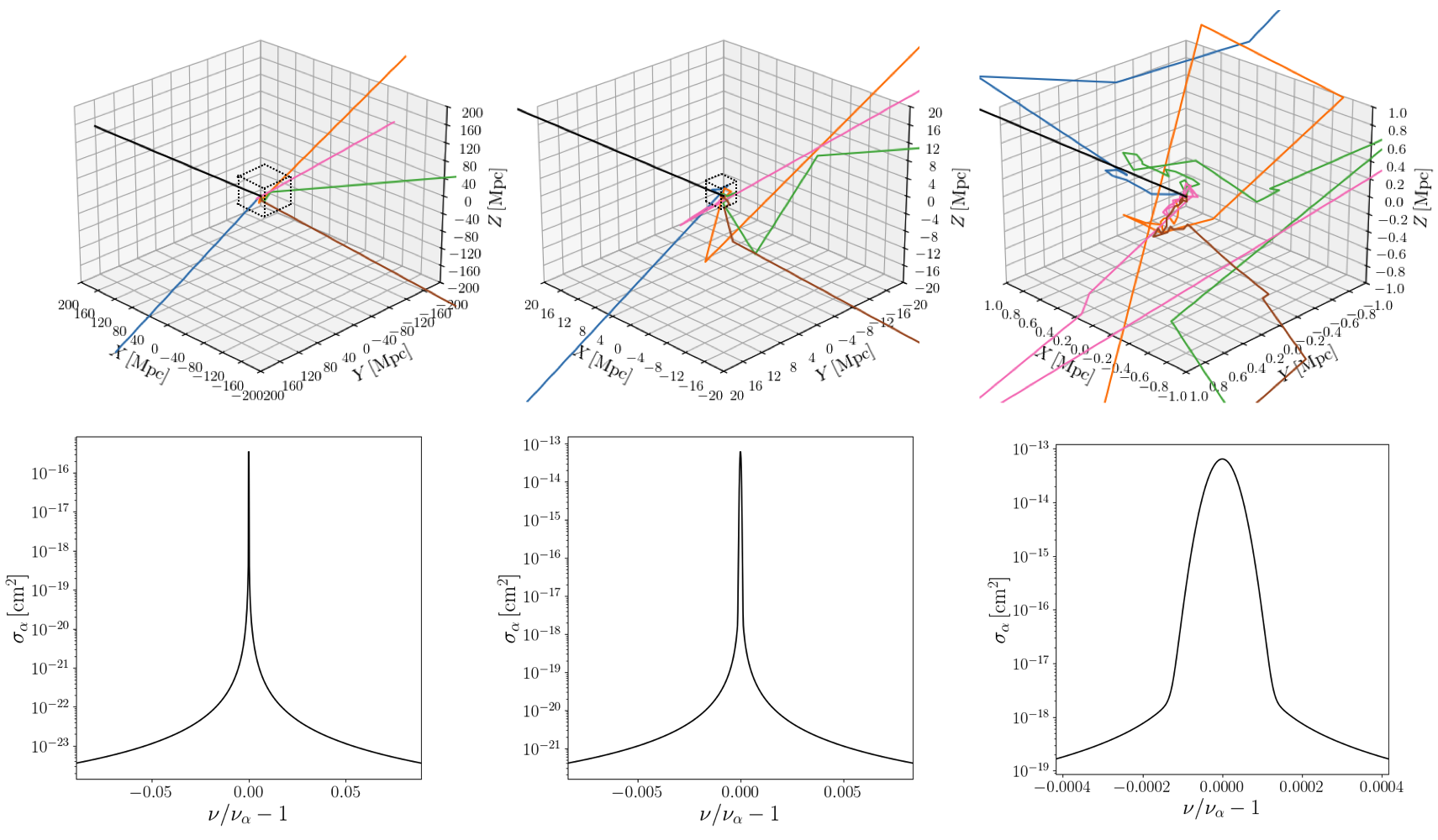}
\caption{\textbf{Top row}: trajectories of five photons, as was simulated by {\tt SP$\alpha$RTA}. All photons are absorbed at the center once their apparent frequency reaches Ly$\alpha$ at $z_\mathrm{abs}=10$. It is assumed that the photons travel in a homogeneous IGM with $T_k=10^4\,\mathrm{K}$ and $x_\mathrm{HI}=1$ (though the gas is allowed to move with some peculiar velocity with respect to the comoving frame). For reference, we show by the black line the trajectory of a photon propagating in a straight-line. In all panels the same trajectories are shown, albeit on different scales (from left to right, each panel shows a progressively zoomed-in view of the sub-volume highlighted in the panel to its left). \textbf{Bottom row}: Voigt cross-section, Eq.~\eqref{eq: 14}, as seen in different spectral resolutions (the spectral boundaries at the bottom row match the spatial boundaries at the top row). Zooming-in to smaller scales (left to right panels), the approximation of the cross-section as a delta function becomes less valid, increasing the importance of multiple scatterings.}
\label{fig: 1}
\end{figure*}
\begin{figure*}
\begin{centering}
\includegraphics[width=0.9\textwidth]{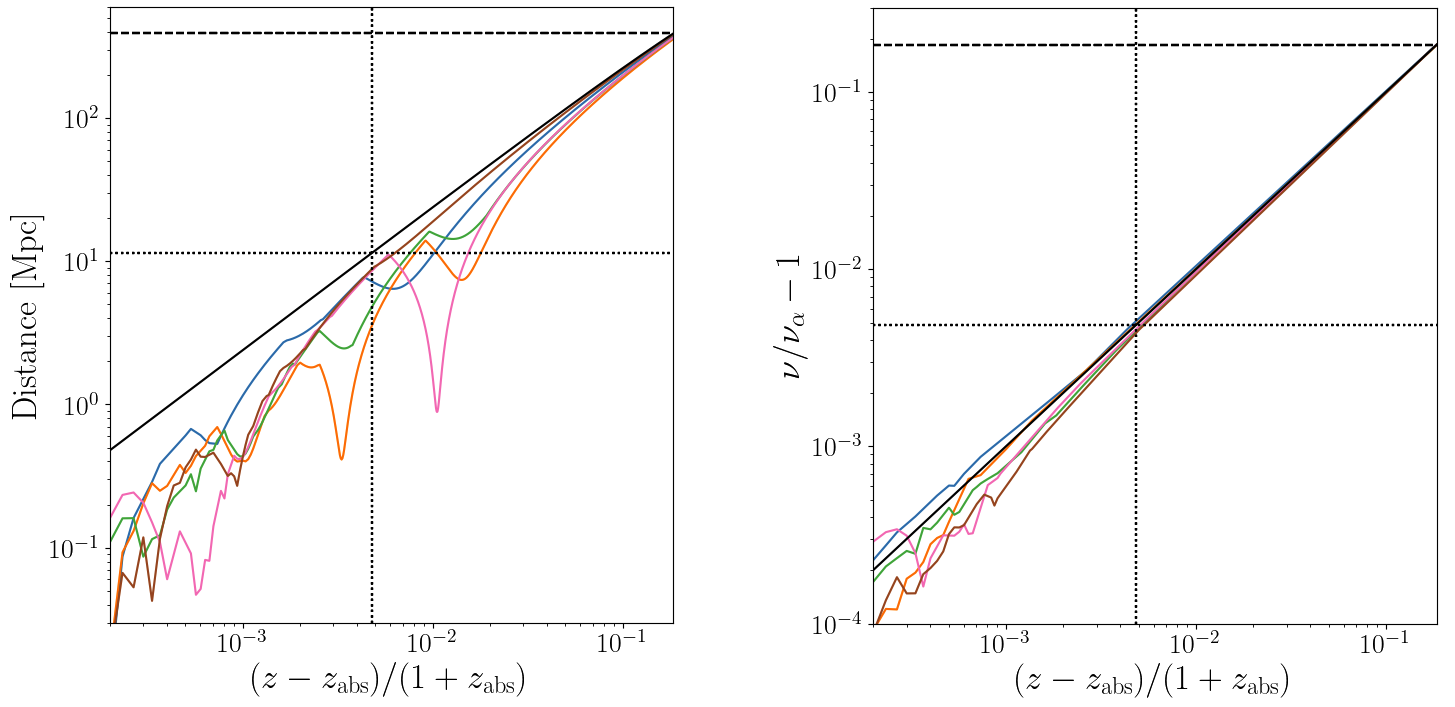}
\caption{\textbf{Left panel}: Radial distance of the photons from the absorption point (at $z_\mathrm{abs}=10$) as a function of redshift. \textbf{Right panel}: Apparent frequency shift of the photons from Ly$\alpha$ as a function of redshift. The colors of the curves here match the colors of the trajectories in Fig.~\ref{fig: 1}. Specifically, the black curve shown here corresponds to a photon that propagates in a straight-line with apparent frequency that scales with redshift, $\nu_\mathrm{app}\propto\left(1+z\right)$. The horizontal and vertical dotted lines correspond to the diffusion scale (or frequency), as given by Eqs.~\eqref{eq: 23}-\eqref{eq: 24}. The horizontal dashed black line corresponds to the end of the simulation, when the apparent frequency of the photons reached Ly$\beta$.}
\label{fig: 2}
\end{centering}
\end{figure*}

To gain some intuition for the expected behavior of the desired MS window function, we show in Fig.~\ref{fig: 1} the trajectories of five photons, as simulated by {\tt SP$\alpha$RTA}. In all panels, all photons are absorbed at the center at $z_\mathrm{abs}=10$. Interestingly, on large scales (left panel), all photons appear to travel in straight-lines, while on small scales, they appear to perform a random-walk with a decreasing step-size as the scale becomes smaller. This makes sense as on large scales the apparent frequency of the photons is much higher than $\nu_\alpha$, where the cross-section and the optical depth are sufficiently small. Evidently, on large scales the damping wing is less visible, making the cross-section to appear as a delta function. In contrast, on small scales, as the apparent frequency of the photons approaches Ly$\alpha$, the mean-free-path decreases significantly due to the large cross-section.

The above features can also be seen in Fig.~\ref{fig: 2}, where we show in the left panel the radial distance between the photons from Fig.~\ref{fig: 1} and the absorption point as a function of redshift. For reference, the black curve in that plot corresponds to a photon propagating in a straight-line. Indeed, as expected, the radial distance in all curves is shorter than the straight-line distance, $r<R_\mathrm{SL}$. Furthermore, on large scales, the radial distance of all the photons from the origin approaches the straight-line limit, $r\to R_\mathrm{SL}$. In contrast, on small scales, we can see that the radial distance becomes shorter compared to $R_\mathrm{SL}$, and that trend continues as we go to smaller scales. Note that the sharp cusps do not indicate a scattering event (unlike the local maxima in that plot), but rather ``closest passage'' points of the photons with respect to the origin.

On the right panel of Fig.~\ref{fig: 2} we present the apparent frequency of the photons, $\nu_\mathrm{app}$, as a function of redshift. Here, the black curve corresponds to a photon with an apparent frequency that scales with redshift, $\nu_\mathrm{app}\propto\left(1+z\right)$. We see that on small scales, the apparent frequency is modified by small relative peculiar and thermal velocities of order $v_\mathrm{rel}/c\sim10^{-5}$. However, on large scales, the effect of the relative velocities becomes negligible compared to redshift.

\newpage

\section{Analytical window function}\label{sec: Analytical window function}

\subsection{Diffusion scale}\label{subsec: Diffusion scale}

\begin{figure*}
\begin{centering}
\includegraphics[width=\textwidth]{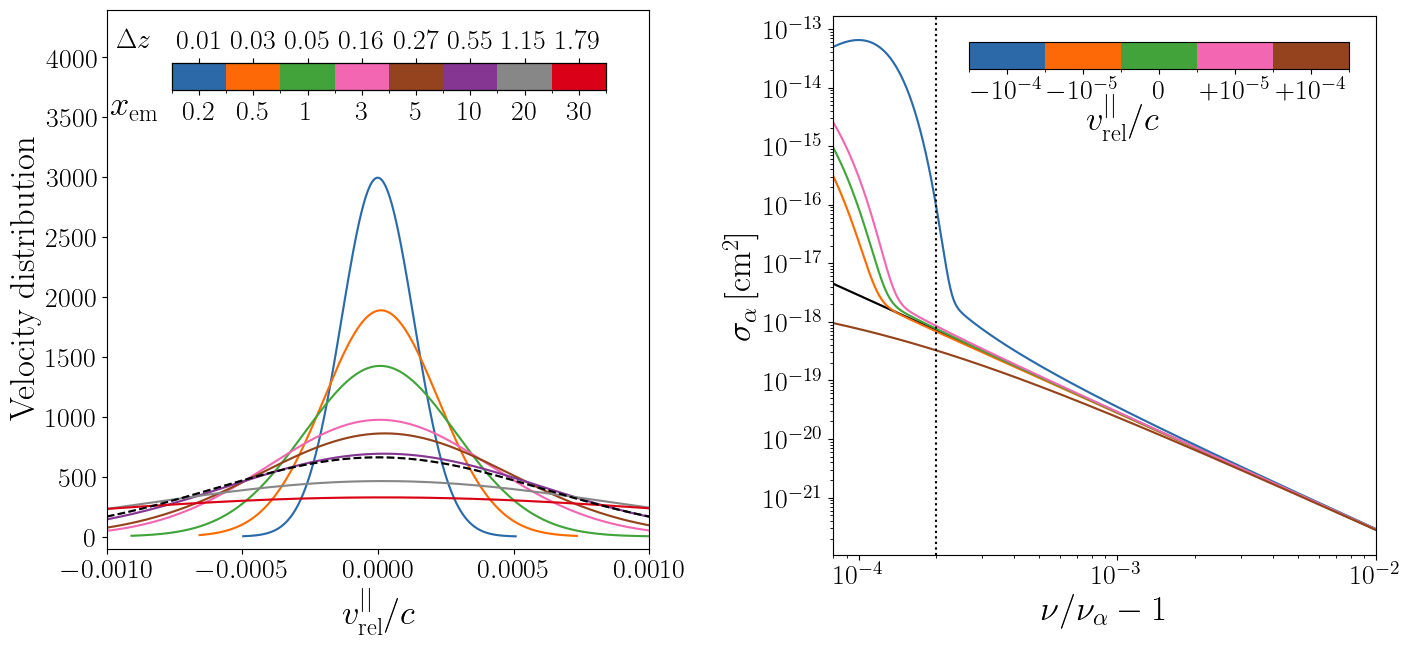}
\caption{\textbf{Left panel}: distribution of the relative peculiar velocity between the absorber at $z_\mathrm{abs}=10$ and the emitter (whose redshift is implied by the $x_\mathrm{em}$ value), as extracted from the {\tt SP$\alpha$RTA} simulation. All curves are cut at the highest and lowest values obtained in the simulation. We also show, by the dashed black curve, the Gaussian distribution that corresponds to $|\mathbf v|/c$ at $z=0$ (computed in Newtonian gauge), smoothed on a scale of $10\,\mathrm{Mpc}$. \textbf{Right panel}: the apparent Voigt cross-section $\sigma_\alpha\left(\nu,T_k\right)$ (Eq.~\ref{eq: 14}) as modified by relative peculiar velocities due to Doppler shift. In all curves (except the black curve) $T_k=10^4\,\mathrm{K}$. The black curve corresponds to $T_k=0$ and zero relative peculiar velocity. The vertical dotted line corresponds to the mean initial frequency in the {\tt SP$\alpha$RTA} simulation (c.f. right panel of Fig.~\ref{fig: 2}).}  
\label{fig: 4}
\end{centering}
\end{figure*}

The scale that separates ``small'' scales from ``large'' scales is the diffusion scale. In order to analytically estimate it, we follow the steps done by Ref.~\cite{Loeb:1999er}, though we solve for the comoving diffusion scale, not the proper one. We thus seek, for a given absorption redshift $z_\mathrm{abs}$, the redshift $z_*$ for which a photon that comes from infinity through a homogeneous IGM will experience an optical depth of unity,
\begin{equation}\label{eq: 21}
\tau_*=\int_{z_{*}}^{\infty}\frac{cn_{\mathrm{HI}}\left(z'\right)\sigma_{\alpha}\left(\nu';T_k\right)}{H\left(z'\right)\left(1+z'\right)}dz'\overset{!}{=}1.
\end{equation}
The logic behind this definition is that for $z>z_*$ the optical depth is sufficiently small that it is unlikely for a scattering event to occur, thereby above this redshift photons travel approximately in straight-lines. To proceed with the derivation, we shall assume matter-domination, $H\left(z\right)\simeq H_0\left(1+z\right)^{3/2}$, and assume that the apparent frequency of the photon only redshifts (with no Doppler shift), $\nu'=\left(1+z'\right)/\left(1+z_\mathrm{abs}\right)\nu_\alpha$. As for the cross-section, we shall take the wing approximation where $H\left(a_T;x_T\right)\approx a_T/\left(\pi^{1/2}x_T^2\right)$, resulting in $\sigma_{\alpha}\left(\nu';T_k\right)\approx 3\lambda_\alpha^2A_\alpha^2/\left[32\pi^{3}\left(\nu'-\nu_\alpha\right)^2\right]$. The last approximation we shall do would be to ignore all the redshift evolution in the integrand, substituting $z'=z_\mathrm{abs}$, besides the one in $\nu'$. This may seem like a very crude approximation, however it can be shown that the largest contribution to the integral comes from $z_*\approx z_\mathrm{abs}$ (as we shall verify below), where the cross-section is largest. This leaves us with a simple integral that can be solved analytically, and to invert the result to solve for\footnote{We note that Eq.~\eqref{eq: 23} is the same as Eq.~(8) in Ref.~\cite{Loeb:1999er}, when assuming matter-domination in a flat Universe.} $\Delta z_*\equiv z_*-z_\mathrm{abs}$ (or $\Delta \nu_*\equiv \nu_*-\nu_\alpha$),
\begin{flalign}\label{eq: 23}
&\nonumber\frac{\Delta\nu_{*}}{\nu_{\alpha}}=\frac{\Delta z_{*}}{1+z_{\mathrm{abs}}}=\frac{3c^{3}A_{\alpha}^{2}n_{\mathrm{H0}}}{32\pi^{3}\nu_{\alpha}^{4}H_{0}\Omega_{m}^{1/2}}x_{\mathrm{HI}}\left(1+z_{\mathrm{abs}}\right)^{3/2}&
\\&\nonumber\simeq 4.2\times10^{-3}x_{\mathrm{HI}}\left(\frac{\Omega_{b}h^{2}}{0.0223}\right)\left(\frac{\Omega_{m}h^{2}}{0.143}\right)^{-1/2}&
\\&\hspace{20mm}\times\left(\frac{1-Y_{\mathrm{He}}}{0.755}\right)\left(\frac{1+z_{\mathrm{abs}}}{10}\right)^{3/2}.&
\end{flalign}
Notice that this result is in agreement with the wing approximation we took (c.f. right panel of Fig.~\ref{fig: 4}).

Moreover, after plugging Eq.~\eqref{eq: 23} in (the Taylor approximation of) Eq.~\eqref{eq: 7}, one finds the comoving diffusion scale,
\begin{flalign}\label{eq: 24}
&\nonumber R_{*}\left(z_\mathrm{abs}\right)\equiv R_\mathrm{SL}\left(z_\mathrm{abs},z_*\right)\approx\frac{3c^{4}A_{\alpha}^{2}n_{\mathrm{H0}}}{32\pi^{3}\nu_{\alpha}^{4}H_{0}^{2}\Omega_{m}}x_{\mathrm{HI}}\left(1+z_{\mathrm{abs}}\right)&
\\&\nonumber\simeq10.4\,x_{\mathrm{HI}}\left(\frac{\Omega_{b}h^{2}}{0.0223}\right)\left(\frac{\Omega_{m}h^{2}}{0.143}\right)^{-1}&
\\&\hspace{15mm}\times\left(\frac{1-Y_{\mathrm{He}}}{0.755}\right)\left(\frac{1+z_{\mathrm{abs}}}{10}\right)\,\mathrm{Mpc}.&
\end{flalign}
This result agrees with the trajectories we show in Figs.~\ref{fig: 1} and ~\ref{fig: 2}. Photons that come from $R_\mathrm{SL}\gtrsim R_*$ appear to travel in straight-lines, while photons that come from $R_\mathrm{SL}\lesssim R_*$ tend to diffuse and scatter many times before they are absorbed by the IGM. It is therefore convenient to define the following dimensionless parameter,
\begin{equation}\label{eq: 25}
x_\mathrm{em}\equiv\frac{R_\mathrm{SL}\left(z_\mathrm{abs},z_\mathrm{em}\right)}{R_{*}\left(z_\mathrm{abs}\right)}.
\end{equation}

When $x_\mathrm{em}\gg1$, we can thus expect that photons will travel mostly in straight-lines, before their apparent frequency reaches Ly$\alpha$. If we consider the radial distribution $f_{R_\mathrm{SL}}\left(r\,;z\right)$ from Sec.~\ref{subsec: Window function}, we may say that in the limit $x_\mathrm{em}\to\infty$, $f_{R_\mathrm{SL}}\left(r\,;z\right)\to\delta^\mathrm{D}\left(r-R_\mathrm{SL}\right)$, or $f_1\left(y;z\right)\to\delta^\mathrm{D}\left(y-1\right)$. This must be true, because when photons traverse infinitely large distances without being scattered, this straight-line segment is the dominant component in the photons trajectory.

What about the other limit, when $x_\mathrm{em}\to0$? From Fig.~\ref{fig: 2} we can see that as the scale becomes smaller and frequency of the photon approaches Ly$\alpha$, its effective distance from the origin becomes shorther, compared to $R_\mathrm{SL}$. In fact, when $T_k=0$, it can be shown analytically that this limit corresponds to a radial distribution of $f_1\left(y;z\right)\to\delta^\mathrm{D}\left(y\right)$. To see that, consider the analytical distribution of Ref.~\cite{Loeb:1999er} in the diffusion limit. For the radial coordinate $r$, this distribution is a Maxwellian with a sigma parameter given by Eq.~\eqref{eq: 18}. Now, use $\left(\nu_\mathrm{em}-\nu_\alpha\right)/\Delta\nu_*\left(z_\mathrm{abs}\right)\approx R_{\mathrm{SL}}\left(z_{\mathrm{abs}},z_{\mathrm{em}}\right)/R_{*}\left(z_{\mathrm{abs}}\right)=x_\mathrm{em}$ (this approximation works well as long as the correlated velocities on the smallest scales do not change the apparent frequency of the photon) to arrive at $\sigma_\mathbf{x}/R_{\mathrm{SL}}\left(z_{\mathrm{abs}},z_{\mathrm{em}}\right)=\left(2x_\mathrm{em}/9\right)^{1/2}\to0$. This means that the (normalized) distribution of $y\equiv r/R_\mathrm{SL}$ has zero mean and RMS, implying that $y$ distributes as $\delta^\mathrm{D}\left(y\right)$. For a finite $T_k$ however, it can be shown, using the distribution of Ref.~\cite{Smith:2025jsc}, that when $x_T\to0$, the mean and RMS of the radial coordinate $r$ are proportional to $v_\mathrm{th}/H\left(z_\mathrm{abs}\right)$. In other words, there is a sphere of radius $v_\mathrm{th}/H\left(z_\mathrm{abs}\right)\sim100\left(T_k/10^4\,\mathrm{K}\right)^{1/2}\,\mathrm{kpc}$ around the point of absorption, where Ly$\alpha$ photons can be found. This conclusion can be physically understood, as the random thermal velocities of the hydrogen atoms shift constantly the apparent frequency of the scattered photon blueward and redward with respect to the line center. Hence, it is meaningless to consider the distribution of the dimensionless variable $y$ when $x_\mathrm{em}\to0$ for a finite $T_k$. Still, since we are not interested in sub-Mpc scales in our {\tt 21cmFAST} simulation, we can expect that for the relevant scales, the distribution of 
$y$ will favor lower values when $x_\mathrm{em}\ll1$.

Since the distribution of $y$ for $x_\mathrm{em}\gg1$ is expected to approach $\delta^\mathrm{D}\left(y-1\right)$, while for $x_\mathrm{em}\ll1$ it is expected to approach $\delta^\mathrm{D}\left(y\right)$, we can guess that for $x_\mathrm{em}\simeq1$ the distribution of $y$ will be symmetrical around $y=1/2$ (or equivalently, the distribution of $r$ will be symmetric around $r=R_\mathrm{SL}/2$). In other words, based on Eq.~\eqref{eq: 24}, we can already predict that for $z_\mathrm{abs}\simeq10$, most of the photons that are emitted at a redshift that corresponds to $R_\mathrm{SL}\left(z_\mathrm{abs},z_\mathrm{em}\right)\sim R_*\left(z_\mathrm{abs}\right)\simeq10\,\mathrm{Mpc}$, will only cover a radial distance of $5\,\mathrm{Mpc}$ before they are absorbed!

It is quite remarkable that the simple analytic approach of Ref.~\cite{Loeb:1999er} manages to capture the characteristic scale that separates ``large'' scales from ``small'' scales, even though in our analysis we have considered more advance physical features, including peculiar velocity and finite temperature. We will see next why it is the case.

\begin{figure}
\begin{centering}
\includegraphics[width=\columnwidth]{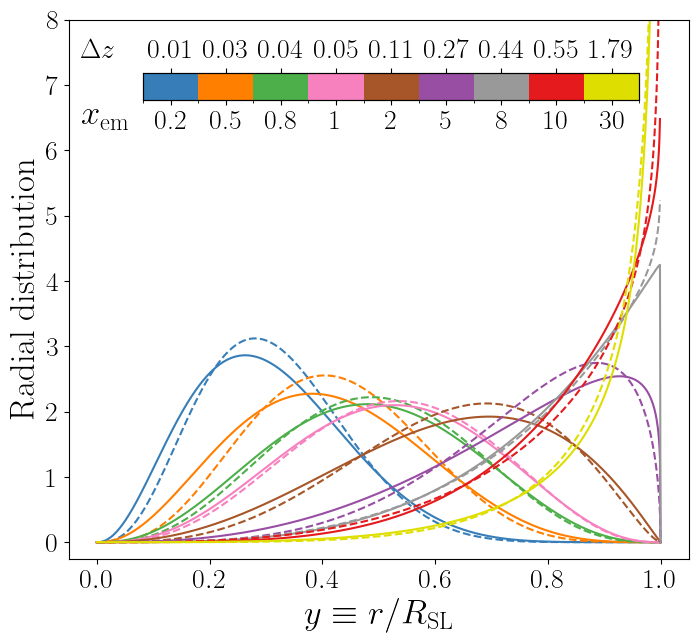}
\caption{Radial distributions $f_\mathrm{MS}\left(y;z_\mathrm{abs}\right)$ of the normalized distance variable, $y\equiv r/R_\mathrm{SL}$, for different values of $x_\mathrm{em}\equiv R_\mathrm{SL}\left(z_\mathrm{abs},z_\mathrm{em}\right)/R_*\left(z_\mathrm{abs}\right)$ (for better interpretation of that quantity, we report also the corresponding $\Delta z = z_\mathrm{em}-z_\mathrm{abs}$). All curves shown here were obtained by fitting the numerical distributions from {\tt SP$\alpha$RTA} to a beta function. \emph{Solid (dashed)} curves correspond to having turned on (off) all the physical features listed in Sec.~\ref{sec: SPaRTA}. All distributions correspond to $z_\mathrm{abs}=10$, $T_k=10^4\,\mathrm{K}$ and $x_\mathrm{HI}=1$.} 
\label{fig: 3}
\end{centering}
\end{figure}

\begin{figure}
\begin{centering}
\includegraphics[width=\columnwidth]{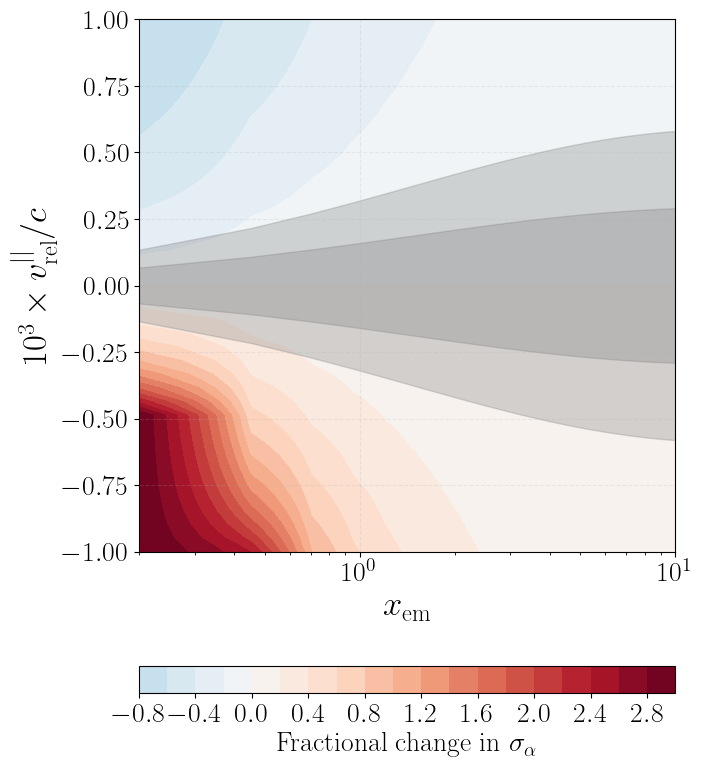}
\caption{Fractional change in the Ly$\alpha$ cross-section (Eq.~\ref{eq: 14}), due to relative velocities. The cross-section is computed at $T_k=10^4\,\mathrm{K}$ at an apparent frequency $\nu_\mathrm{app}=\nu_\alpha\left(1+z_\mathrm{em}\right)/\left(1+z_\mathrm{abs}\right)/\left(1-v_\mathrm{rel}^{||}/c\right)$, with $z_\mathrm{abs}=10$ and $z_\mathrm{em}$ is determined from $z_\mathrm{abs}$ and $x_\mathrm{em}$ (see Eq.~\ref{eq: 25}). Note that the cross-section diverges when the relative velocity counters the Hubble flow, i.e. when $v_\mathrm{rel}^{||}/c\approx -\left(z_\mathrm{em}-z_\mathrm{abs}\right)/\left(1+z_\mathrm{abs}\right)$, but for better visibility we clip the maximal fractional change at 3. Grey shaded regions mark the 1$\sigma$ and 2$\sigma$ of the relative velocity distribution (c.f.~Fig.~\ref{fig: 4}).}
\label{fig: 5}
\end{centering}
\end{figure}

\subsection{Beta distributions}\label{subsec: Beta distributions}

We look for an analytical distribution that would describe the various radial distributions. This distribution has a finite support, between $[0,1]$ (for the dimensionless variable $y\equiv r/R_\mathrm{SL}$), and should have a single peak that can be found anywhere in that range. The beta distribution is an analytical distribution that satisfies all of these properties. It is given by
\begin{equation}\label{eq: 26}
f_\mathrm{MS}\left(y\right)=\frac{y^{\alpha-1}\left(1-y\right)^{\beta-1}}{\mathrm B\left(\alpha,\beta\right)},
\end{equation}
where $\mathrm B\left(\alpha,\beta\right)$ is the beta function, and $\alpha,\,\beta>0$ are the parameters of the distribution. The mean $\mu$ and the variance $\sigma^2$ of the beta distribution are
\begin{equation}\label{eq: 27}
\mu=\frac{\alpha}{\alpha+\beta},\qquad\sigma^2=\frac{\alpha\beta}{\left(\alpha+\beta\right)^2\left(\alpha+\beta+1\right)}.
\end{equation}
If $\alpha\gg\beta$ ($\alpha\ll\beta$), the mean of the distribution goes towards $\mu\to1$ ($\mu\to0$). Interestingly, in the limit $\alpha/\beta\to\infty$ ($\alpha/\beta\to0$), the beta distribution goes to $f_\mathrm{MS}\left(y\right)\to\delta^\mathrm{D}\left(y-1\right)$ ($f_\mathrm{MS}\left(y\right)\to\delta^\mathrm{D}\left(y\right)$), namely this is the SL (diffusion) limit. This can be understood from the fact that the only normalized distribution with a vanishing variance is a Dirac delta distribution, centered at the mean value. Based on the discussion above, we can therefore predict that for our application, the radial distribution for $x_\mathrm{em}\gg1$ ($x_\mathrm{em}\ll1$) corresponds to a beta distribution with $\alpha\gg\beta$ ($\alpha\ll\beta$), while $\alpha\sim\beta$ for $x_\mathrm{em}\sim1$.

We thus choose to model the MS radial distribution as a beta distribution. To find $\alpha$ and $\beta$, we apply the following procedure. After running an MC simulation with {\tt SP$\alpha$RTA}, we are left with a 2D dataset, whose one axis corresponds to the normalized radial distance $y\equiv r/R_\mathrm{SL}$, whereas the second axis corresponds to the redshift of the emitter $z_\mathrm{em}$, which, given the absorption redshift $z_\mathrm{abs}$ of the simulation, is equivalent to having $x_\mathrm{em}\equiv R_\mathrm{SL}\left(z_\mathrm{abs},z_\mathrm{em}\right)/R_*\left(z_\mathrm{abs}\right)$ (we remind the reader that each intermediate point in the {\tt SP$\alpha$RTA} simulation can be considered as a possible source of emission). We then split the data along the second axis into bins of width $\Delta x_\mathrm{em}=0.1$. Then, for a given $x_\mathrm{em}$ value, we select only the data points that fall into the corresponding bin. Once the selected data points (that contain only the $y$ values) are at hand, we compute their sample mean and variance, and upon inverting the relations in Eq.~\eqref{eq: 27}, the parameters $\alpha$ and $\beta$ are inferred. This is the most straightforward method to fit the 1D data to a beta distribution, without assessing the goodness of the fit. We have confirmed however, by binning the $y$ axis of the 2D data and fitting directly the resulting histogram, that the distribution of the data is indeed similar to a beta distribution, given an $x_\mathrm{em}$ value.

We present in Fig.~\ref{fig: 3} the radial distributions $f_\mathrm{MS}\left(y;z_\mathrm{abs}\right)$ for $z_\mathrm{abs}=10$. As expected, distributions with high (low) values of $x_\mathrm{em}$ prefer high (low) values of $y$. We can also see how the beta distribution approaches (slowly) $\delta^\mathrm{D}\left(y-1\right)$ when $x_\mathrm{em}\to\infty$. In addition, we see that the distribution is more symmetrical around $y=1/2$ for $x_\mathrm{em}=0.8$ rather than $x_\mathrm{em}=1$, implying that $0.8R_*\left(z_\mathrm{abs}\right)$ is more appropriate to be the scale that separates ``large'' scales from ``small'' scales.

The solid (dashed) curves in Fig.~\ref{fig: 3} correspond to having turned on (off) all the physical features listed in Sec.~\ref{sec: SPaRTA}: peculiar velocity, finite temperature, anisotropic scattering and atom recoil. Interestingly, we see that these physical features do not affect considerably the radial distributions, and these distributions are mostly determined by the $x_\mathrm{em}$ value. To our surprise, we find that peculiar velocities and finite temperature have negligible effects on the radial distributions. In fact, we find that the physical feature that has the most significant effect in determining the shape of the radial distribution is anisotropic scattering (Eq.~\ref{eq: 16}). Nevertheless, it is important to stress that while the radial distributions are not sensitive to the effect of peculiar velocity nor finite temperature, the \emph{realizations} of the photon trajectories are sensitive to these effects --- only when gathering enough statistics the differences caused by these effects become small.    

From the right panel of Fig.~\ref{fig: 4} it is easy to see why the radial distributions of {\tt SP$\alpha$RTA} do not depend on the temperature of the IGM (at least for $T_k\lesssim10^4\,\mathrm{K}$). Comparison between the cross-section at $T_k=0$ (black curve) and the cross-section at $T_k=10^4\,\mathrm{K}$ (green curve) shows that the difference between them begins at the core, at $\Delta\nu/\nu_\alpha\lesssim2\times10^{-4}$. However, this value corresponds to the mean initial frequency in the {\tt SP$\alpha$RTA} simulation (c.f. right panel of Fig.~\ref{fig: 2}). Of course, had we studied the radial distributions on much smaller scales, where the analytical formula of Ref.~\cite{Loeb:1999er} fails, then the thermal motions of the hydrogen atoms would alter significantly the apparent frequency of the photons when it is close to the line center, thereby possibly changing the radial distributions on these scales.  

To understand why the peculiar velocities do not alter much the radial distributions, the left panel of Fig.~\ref{fig: 4} shows that the parallel component of the relative peculiar velocity distributes as a Gaussian with mean zero and variance that becomes wider with higher values of $x_\mathrm{em}$. This is not surprising, since we have assumed in {\tt SP$\alpha$RTA} that the peculiar velocity field distributes as a Gaussian according to linear perturbation theory. If we denote by $\sigma^2\left(z\right)\equiv \left\langle\left(v^{||}\left(z\right)\right)^2\right\rangle$ the variance of the parallel component of the peculiar velocity at some redshift $z$, then the variance of the parallel component of the relative velocity is given by $\left\langle\left(v_\mathrm{rel}^{||}\left(z_\mathrm{abs},z_\mathrm{em}\right)\right)^2\right\rangle=\sigma^2\left(z_\mathrm{abs}\right)+\sigma^2\left(z_\mathrm{em}\right)-2\rho^{||}\left(z_\mathrm{abs},z_\mathrm{em}\right)\sigma\left(z_\mathrm{abs}\right)\sigma\left(z_\mathrm{em}\right)$, where $\rho^{||}\left(z_\mathrm{abs},z_\mathrm{em}\right)$ is the Pearson correlation coefficient (see exact definition at Appendix \ref{sec: Conditional Gaussian random variables}). If $z_\mathrm{em}=z_\mathrm{abs}+\Delta z$, then $\rho^{||}\left(z_\mathrm{abs},z_\mathrm{em}\right)=1-\mathcal O\left(\Delta z\right)$, and for small $\Delta z$ the variance of the relative velocity becomes proportional to $\Delta z$. For example, at $x_\mathrm{em}=0.2$ the RMS of $v_\mathrm{rel}^{||}/c$ is of order $\mathcal O\left(10^{-5}\right)$. From the right panel of Fig.~\ref{fig: 4} we see that relative velocities of this size barely change the effective cross-section for the frequency shifts considered in this work. On the other hand, while relative velocities of order $\mathcal O\left(10^{-4}\right)$ do have a more dramatic effect on the effective cross-section, such relative velocities correspond to scales of $x_\mathrm{em}\gtrsim 1$, where the difference in the cross-section becomes small and there are far fewer scattering events. This conclusion is also supported by Fig.~\ref{fig: 5}, where the fractional change in the cross-section is displayed, as a function of the relative velocity and $x_\mathrm{em}$; as indicated by the grey regions, it is very unlikely to have a relative velocity large enough to significantly alter the apparent cross-section that the photon sees, even on scales as small as $x_\mathrm{em}=0.2$.

\begin{figure}
\begin{centering}
\includegraphics[width=\columnwidth]{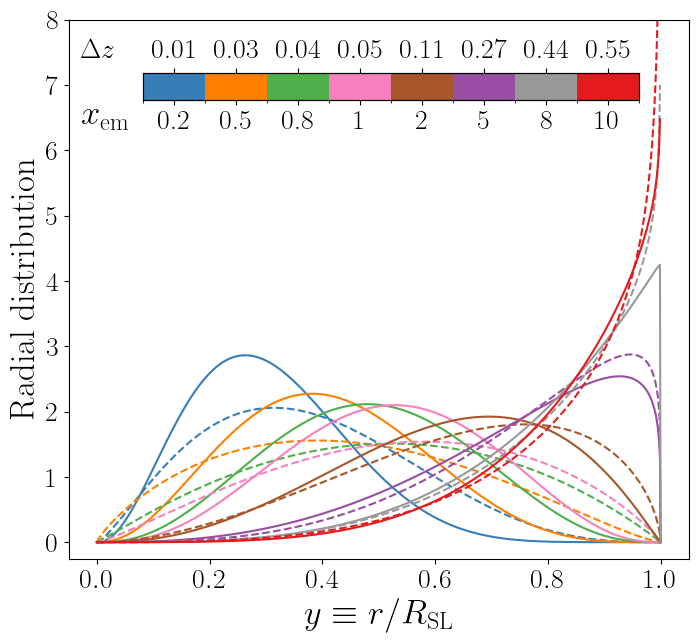}
\caption{Similarly as Fig.~\ref{fig: 3}, but here solid curves correspond to accounting for correlations in the peculiar velocity field, while the dashed curves correspond to zeroing the Pearson correlation coefficient $\rho$ (see exact definition at Appendix \ref{sec: Conditional Gaussian random variables}), meaning that all velocity samples in the simulation are completely uncorrelated.} 
\label{fig: 6}
\end{centering}
\end{figure}

Our conclusion that peculiar velocities have little effect on the radial distributions (and hence on $J_\alpha$ maps as well when the effect of Ly$\alpha$ MS is studied) is in contrast with the findings of Refs.~\cite{Semelin:2023lhz, Mittal:2023xih}. Unlike this work, these works considered non-linear corrections to the peculiar velocity field. It is possible that the absolute value of the relative peculiar velocity could be enhanced on small scales due to non-linear effects. While the study of non-linear effects on the radial distributions is beyond the scope of this work\footnote{But see Appendix \ref{sec: Beyond linear perturbation theory} where we validate our results against a 2LPT simulation.}, we show the importance of the peculiar velocity correlations (in linear theory) in Fig.~\ref{fig: 6}. In that figure, the dashed curves correspond to setting $\rho\left(z_\mathrm{abs},z_\mathrm{em}\right)=0$, thus the randomly drawn velocity vector in our simulation is completely uncorrelated with past samples. Indeed, the velocity correlation is very important on small scales, while the radial distributions on large scales are unaffected by the correlations in the velocity field. We leave a more thorough study of non-linear effects on our semi-analytic formalism to future work.

\begin{figure}[t]
\begin{centering}
\includegraphics[width=\columnwidth]{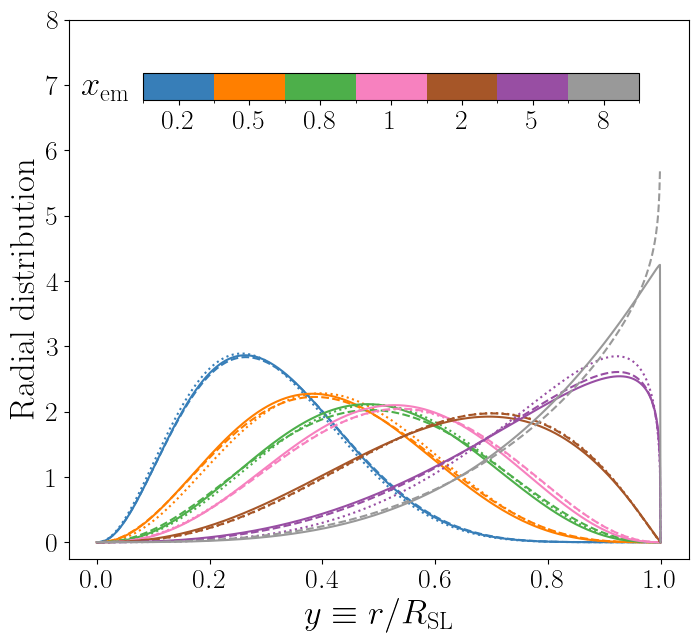}
\caption{Similarly as Fig.~\ref{fig: 3}, but here solid, dashed and dotted curves correspond to $z_\mathrm{abs}=10$, $z_\mathrm{abs}=20$ and $z_\mathrm{abs}=30$, respectively. Some high $x_\mathrm{em}$ values are not presented for some of the curves, as these $x_\mathrm{em}$ values correspond to scales where the apparent frequency of the photon is higher than Ly$\beta$ (since $x_\mathrm{em}\equiv R_\mathrm{SL}/R_*\left(z_\mathrm{abs}\right)$ and $R_*\left(z_\mathrm{abs}\right)\propto1+z_\mathrm{abs}$, $R_\mathrm{SL}$ ought to be increased if $z_\mathrm{abs}$ is increased while $x_\mathrm{em}$ is fixed, and this consequently increases the frequency of the photon at $z_\mathrm{em}$).}
\label{fig: 7}
\end{centering}
\end{figure}

\begin{figure}
\begin{centering}
\includegraphics[width=\columnwidth]{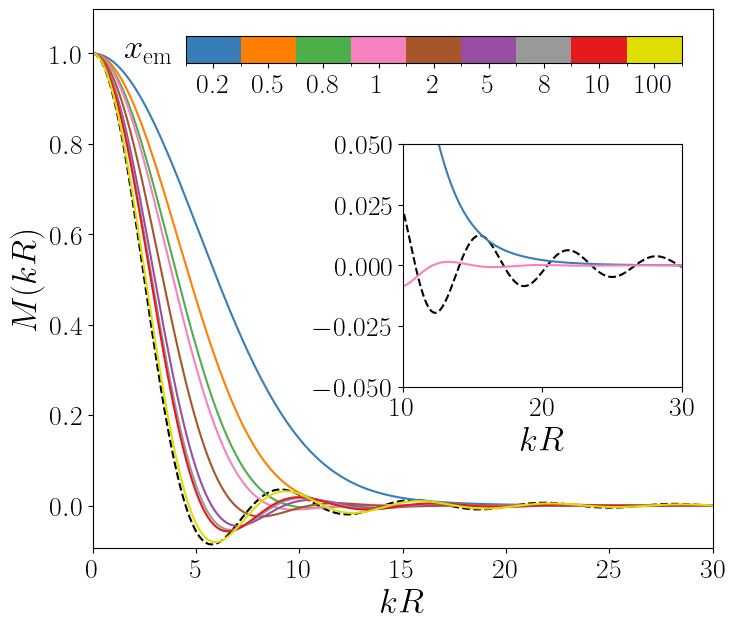}
\caption{The MS window function, as given by Eq.~\eqref{eq: 32} and the fit of Eqs.~\eqref{eq: 28}-\eqref{eq: 30}, for different $x_\mathrm{em}$ values. For comparison, we show the SL window function, Eq.~\eqref{eq: 13}, by the black dashed curve. For better visualization, the inset shows how the window functions behave at high $kR$ values.}
\label{fig: 8}
\end{centering}
\end{figure}

\subsection{Analytical fit}\label{subsec: Analytical fit}

Naively, under the assumption that the fluctuations in $x_\mathrm{HI}$ and $T_k$ are small, one would assume that the radial distributions depend on four parameters: the absorption redshift $z_\mathrm{abs}$, the emission redshift $z_\mathrm{em}$, the IGM ionization fraction $x_\mathrm{HI}$, and the IGM temperature $T_k$. We previously discussed why $T_k$ does not affect the radial distributions for the scales of our interest. That leaves us with three parameters, $z_\mathrm{abs}$, $z_\mathrm{em}$ and $x_\mathrm{HI}$. Note that all these three numbers are encoded in the $x_\mathrm{em}$ parameter, which we have shown to have a major significance in determining the radial distributions, so in theory one of the three parameters can be replaced with $x_\mathrm{em}$. However, we have confirmed that for a given $x_\mathrm{em}$, the radial distributions do not depend significantly on $x_\mathrm{HI}$, i.e. the $x_\mathrm{HI}$ information is well characterized by $x_\mathrm{em}$, given $z_\mathrm{abs}$ and $z_\mathrm{em}$. That leaves us with only two parameters, $x_\mathrm{em}$ and $z_\mathrm{abs}$. Yet, given $x_\mathrm{em}$, it appears that the radial distributions do not depend strongly on $z_\mathrm{abs}$, as we show in Fig.~\ref{fig: 7}. Therefore, we conclude that all the radial distributions are approximately controlled by a single number, $x_\mathrm{em}$.

Instead of fitting directly the $\alpha$ and $\beta$ parameters of the beta distribution, we fit to the mean of the distribution $\mu$ (given in Eq.~\ref{eq: 27}) and $\eta\equiv\alpha/\left(\alpha+\beta^2\right)$. Given these variables, $\alpha$ and $\beta$ are found via
\begin{equation}\label{eq: 28}
\alpha\left(x_\mathrm{em}\right)=\frac{\eta^{-1}\left(x_\mathrm{em}\right)-1}{\left(\mu^{-1}\left(x_\mathrm{em}\right)-1\right)^2},\quad \beta\left(x_\mathrm{em}\right)=\frac{\eta^{-1}\left(x_\mathrm{em}\right)-1}{\mu^{-1}\left(x_\mathrm{em}\right)-1}.
\end{equation}
To find the fit for $\mu\left(x_\mathrm{em}\right)$ and $\eta\left(x_\mathrm{em}\right)$, we have collected all the $\alpha\left(x_\mathrm{em}\right)$ and $\beta\left(x_\mathrm{em}\right)$ samples at the $x_\mathrm{em}$ bins described in Sec.~\ref{subsec: Beta distributions}, which were later converted to $\mu\left(x_\mathrm{em}\right)$ and $\eta\left(x_\mathrm{em}\right)$ data samples. This process was done for $z_\mathrm{abs}=10,\,15,\,20,\,25$ and $30$. We then found the following fit, which matches the data points at all these redshifts within $10\%$ residuals,

\begin{flalign}\label{eq: 29}
&\nonumber\mu\left(x_{\mathrm{em}}\right)=&
\\&\begin{cases}
0.3982x_{\mathrm{em}}^{0.1592} & x_{\mathrm{em}}\leq0.2\\
\big\{-0.0285\zeta_{\mathrm{em}}^{5}+0.087\zeta_{\mathrm{em}}^{4}-0.1205\zeta_{\mathrm{em}}^{3}\\-0.0456\zeta_{\mathrm{em}}^{2}+0.3787\zeta_{\mathrm{em}}+0.5285\big\} & 0.2<x_{\mathrm{em}}\leq3\\
\big\{-0.104\zeta_{\mathrm{em}}^{5}+0.4867\zeta_{\mathrm{em}}^{4}-0.8217\zeta_{\mathrm{em}}^{3}\\+0.4889\zeta_{\mathrm{em}}^{2}+0.264\zeta_{\mathrm{em}}+0.518\big\} & 3<x_{\mathrm{em}}\leq30\\
1-1.0478x_{\mathrm{em}}^{-0.7266} & 30<x_{\mathrm{em}}
\end{cases}&
\end{flalign}
\begin{flalign}\label{eq: 30}
&\nonumber\eta\left(x_{\mathrm{em}}\right)=&
\\&\begin{cases}
0.4453x_{\mathrm{em}}^{1.296} & x_{\mathrm{em}}\leq0.2\\
\big\{0.352\zeta_{\mathrm{em}}^{5}-0.0516\zeta_{\mathrm{em}}^{4}-0.293\zeta_{\mathrm{em}}^{3}\\+0.342\zeta_{\mathrm{em}}^{2}+0.582\zeta_{\mathrm{em}}+0.266\big\} & 0.2<x_{\mathrm{em}}\leq3\\
\big\{2.17\zeta_{\mathrm{em}}^{5}-8.832\zeta_{\mathrm{em}}^{4}+13.579\zeta_{\mathrm{em}}^{3}\\-10.04\zeta_{\mathrm{em}}^{2}+4.166\zeta_{\mathrm{em}}-0.17\big\} & 3<x_{\mathrm{em}}\leq20\\
1-2.804x_{\mathrm{em}}^{-1.242} & 20<x_{\mathrm{em}}
\end{cases},&
\end{flalign}
where $\zeta_\mathrm{em}\equiv\log_{10}x_\mathrm{em}$.

A few remarks on the fit of Eqs.~\eqref{eq: 29}-\eqref{eq: 30}. Firstly, the fit for $\mu\left(x_{\mathrm{em}}\right)$ is much more accurate compared to $\eta\left(x_{\mathrm{em}}\right)$, and matches the data points from the simulation within sub-percent residuals for most $x_\mathrm{em}$ values and for all $z_\mathrm{abs}$ values we examined. In contrast, the fit to $\eta\left(x_{\mathrm{em}}\right)$ is less accurate as $\eta^{-1}-1$ is a much less numerically stable compared to $\mu^{-1}-1$, especially for large $x_\mathrm{em}$ values. Still, the fit of Eqs.~\eqref{eq: 29}-\eqref{eq: 30} yields a good agreement with the data points at low $x_\mathrm{em}$ values (that are above $x_\mathrm{em}\geq0.2$), which is the important regime for our {\tt 21cmFAST} simulations, as we will see in Sec.~\ref{sec: The effect of Lya multiple scattering on the 21cm signal}. Secondly, while the fit below $x_\mathrm{em}\leq0.2$ is a power-law extrapolation, the fit at the very high $x_\mathrm{em}$ value is not; this fit works well also for $x_\mathrm{em}$ values as high as $x_\mathrm{em}=300$ (we managed to get these $x_\mathrm{em}$ values by simulating an unrealistic IGM with $x_\mathrm{HI}=0.1$ at $z_\mathrm{abs}=10$). According to this fit, when $x_\mathrm{em}\to\infty$, $\mu\to1$ (or equivalently $\alpha/\beta\to\infty$), implying that the radial distribution approaches $\delta^\mathrm{D}\left(r-R_\mathrm{SL}\right)$ in this limit. This confirms our statement from Sec.~\ref{subsec: Diffusion scale}, that the MS window function approaches the SL window function when the scales under consideration are much larger than the diffusion scale. Lastly, we comment that $\mu\left(x_\mathrm{em}\right)=1/2$ when $x_\mathrm{em}\simeq0.85$, this is when $\alpha=\beta$ and the radial distribution becomes symmetrical around $r=R_\mathrm{SL}/2$.

Given our fit, we are finally ready to compute the MS window function. By plugging $f_\mathrm{MS}\left(y\right)$ from Eq.~\eqref{eq: 26} into the expression of the window function, Eq.~\eqref{eq: 8}, one find that
\begin{flalign}\label{eq: 31}
\nonumber&W_\mathrm{MS}\left(kR_\mathrm{SL};z\right)=&
\\&{_2F_{3}}\left(\frac{2+\alpha}{2},\frac{3+\alpha}{2};\frac{3}{2},\frac{2+\alpha+\beta}{2},\frac{3+\alpha+\beta}{2};-\frac{1}{4}k^2R_\mathrm{SL}^2\right).&
\end{flalign}
where $_2F_3$ is a hypergeometric function with $2+3$ parameters, and $\alpha\left(x_\mathrm{em}\right)$ and $\beta\left(x_\mathrm{em}\right)$ are given by our fit, Eqs.~\eqref{eq: 28}-\eqref{eq: 30}. It can be shown that in the limit $x_\mathrm{em}\to\infty$, this window function is nothing more than the simple expression of the SL window function, $W_\mathrm{SL}\left(kR_\mathrm{SL}\right)=\sin kR_\mathrm{SL}/kR_\mathrm{SL}$. This window function however corresponds to an infinitely thin shell of radius $R_\mathrm{SL}$ which is inappropriate for usage in codes that solve numerically the integral of Eq.~\eqref{eq: 3}. Therefore, we ought to find the window function that corresponds to a shell with a finite thickness. According to Eq.~\eqref{eq: 11}, this window function is expressed in terms of $M\left(kR;z\right)$, where $M\left(kR;z\right)$ can be found by integrating over the infinitely thin window function. Plugging $W_\mathrm{MS}\left(kR_\mathrm{SL};z\right)$ from Eq.~\eqref{eq: 31} into the expression of $M\left(kR;z\right)$, Eq.~\eqref{eq: 12}, yields
\begin{flalign}\label{eq: 32}
\nonumber&M_\mathrm{MS}\left(kR;z\right)=&
\\&{_2F_{3}}\left(\frac{2+\alpha}{2},\frac{3+\alpha}{2};\frac{5}{2},\frac{2+\alpha+\beta}{2},\frac{3+\alpha+\beta}{2};-\frac{1}{4}k^2R^2\right).&
\end{flalign}
Again, it is possible to analytically show that when $x_\mathrm{em}\to\infty$, $M_\mathrm{MS}\left(kR;z\right)\to M_\mathrm{SL}\left(kR\right)$, where $M_\mathrm{SL}\left(kR\right)$ is given by Eq.~\eqref{eq: 13}. We show this mathematical property at Fig.~\ref{fig: 8}. We can see that in general the MS window function has more power at low $kR$ values compared to the SL window function, while the opposite is true at high $kR$ values; the MS window function decays faster to zero when $x_\mathrm{em}$ is smaller.

Notice that the redshift-dependence in the RHS of Eq.~\eqref{eq: 32} is encoded within the implicit dependence of $\alpha$ and $\beta$ on $x_\mathrm{em}\equiv R_\mathrm{SL}\left(z,z_\mathrm{em}\right)/R_*\left(z\right)$. The physical reason behind this redshift-dependence is that at higher redshifts the IGM is denser, and the expansion rate of the Universe, via the Hubble parameter, is greater. These two physical effects are combined together to form a diffusion scale which scales as $R_*\left(z\right)\propto1+z$ (c.f.~Eq.~\ref{eq: 24}). This means that for a given $R_\mathrm{SL}$, photons travel a shorter radial distance (with respect to the absorption point) when the absorption redshift $z$ is higher. This redshift-dependence is a necessary feature that is absent from the expression of the SL window function, Eq.~\eqref{eq: 13}.


\section{The impact of Ly$\alpha$ multiple scattering on the 21-cm signal}\label{sec: The effect of Lya multiple scattering on the 21cm signal}

Now that we have an analytical expression for the MS window function, Eq.~\eqref{eq: 32}, we can finally study how Ly$\alpha$ MS affects the 21-cm signal, compared to the SL approximation. We use for that purpose {\tt 21cmFASTv4}~\citep{Davies:2025wsa}, the newest version of {\tt 21cmFAST} that incorporates in the simulation the stochastic nature of the halo field and scatter around astrophysical scaling relations~\citep{Nikolic:2024xxo}. Unlike its predecessors, {\tt 21cmFASTv4} is suitable for implementing our MS window function formalism. In the default settings\footnote{Although the {\tt DexM} halo finder of Ref.~\cite{Mesinger:2007pd} was available since the beginning, its relatively heavy memory requirements made it unsuitable for large parameter studies.} of previous versions of {\tt 21cmFAST}, the emissivity field $\epsilon_*$ was not computed on the grid coordinates, thereby preventing us from filtering it with the MS window function in order to compute $J_\alpha$ (see Eq.~\ref{eq: 3}). This was changed in {\tt 21cmFASTv4}, where the SFRD $\dot\rho_*\left(\mathbf x,z\right)$ is computed on the grid coordinates, while the emissivity field in the code is proportional to $\dot\rho_*\left(\mathbf x,z\right)$. In addition, {\tt 21cmFASTv4} interpolates coeval boxes of the SFRD at different redshifts in order to evaluate the emissivity field at the retarded redshift $z'$, as required by Eqs.~\eqref{eq: 3} and \eqref{eq: 5}. In the astrophysical model that was introduced in {\tt 21cmFASTv4}, above a user-defined mass threshold, the SFRD field receives contributions from discrete halos, each of which has a random star formation rate that is drawn from a log-normal distribution. The integrated contribution to the SFRD field from smaller halos is also taken into account, although it is deterministic given the scaling relations and astrophysical parameters under consideration.  Both atomically and molecularly cooling galaxies contribute to the SFRD, with an average stellar to halo mass relation (SHMR) that is parametrized as a double power-law (c.f. Equation 10 in \cite{Davies:2025wsa}), and a log-normal scatter around this relation.  For a given stellar mass, an individual galaxy's SFR is then determined by sampling from a log-normal conditional probability with a mass and redshift dependent mean and scatter, which characterizes the star forming main sequence (SFMS) of galaxies.  We refer the reader to Ref.~\cite{Davies:2025wsa} for a detailed discussion of the galaxy model.  For simplicity, here we focus on the parameters $f_{\ast,10}$ and $f_{\ast,7}$, corresponding to the normalization of the mean SHMR at halo masses of $10^{10}M_\odot$ and $10^{7}M_\odot$,  for atomic and molecular cooling galaxies, respectively.  This is because the emissivities at all wavelengths scale with these SHMR normalizations.  For simplicity of notation, below we refer to the fiducial choices of these two parameters as  $f_{\rm \ast,fid}^{\rm (II)}$ and $f_{\rm \ast,fid}^{\rm (III)}$, indicating with the superscripts the assumption that PopII and PopIII stars are expected to dominate the stellar populations of atomic and molecular cooling galaxies, respectively.  

For our fiducial settings we adopt the EOS2026 astrophysical parameters~\citep{Breitman2026prep}, which were derived by matching the output of {\tt 21cmFAST} with the UV luminosity function and galaxy angular cross-correlations (as inferred from JWST), as well with reionization history (as inferred from Ly$\alpha$ forest and CMB 
measurements). All the simulations discussed in this section were performed with standard boxes of size $300\,\mathrm{Mpc}$ and a cell size of $1.5\,\mathrm{Mpc}$.

To study the effect of Ly$\alpha$ MS, we have replaced the SL window function in {\tt 21cmFASTv4}, Eq.~\eqref{eq: 13}, with the MS window function, Eq.~\eqref{eq: 32} (we elaborate more on our implementation in Appendix \ref{sec: Implementation of M-MS in 21cmFAST}). This change however is applied only for Ly$\alpha$ photons in the simulation, as X-rays and Lyman Werner photons are still assumed to propagate in straight-lines (namely the fluxes that correspond to these photons are still computed with the SL window function). In order to examine more thoroughly the effect of Ly$\alpha$ MS on the IGM, we have also turned on Ly$\alpha$ heating in the simulation. This heating mechanism, caused by the energy transfer from Ly$\alpha$ photons to the IGM gas, was already incorporated in {\tt 21cmFAST} by Ref.~\cite{Sarkar:2022dvl}, while the implementation followed the same equilibrium equations that can be found in RFB21 and Ref.~\cite{Chen:2003gc} (though see \cite{Raste:2025bpt} for a recent analysis of Ly$\alpha$ heating beyond equilibrium). Since the EOS2026 parameters include $\log_{10}\left(L_X/\mathrm{SFR}\right)=40.5$, where $L_X$ denotes the X-ray luminosity of soft X-ray photons with energy below $2\,\mathrm{keV}$ (in $\mathrm{erg/s}$) and SFR denotes the star formation rate (in $M_\odot/\mathrm{yr}$), the Ly$\alpha$ heating in our fiducial model is sub-dominant to X-ray heating, but it will become more relevant later when we ignore X-ray heating in Sec.~\ref{subsec: Lya heating and Lya multiple scattering}. For these astrophysical settings, the resulting global 21-cm signal has a minimum at $z\simeq14$ and a maximum at $z\simeq7$, implying that Ly$\alpha$ coupling between the spin temperature $T_s$ and the gas kinetic temperature $T_k$ has become efficient already prior to $z\simeq14$.

\begin{figure*}
\begin{centering}
\includegraphics[width=0.95\textwidth]{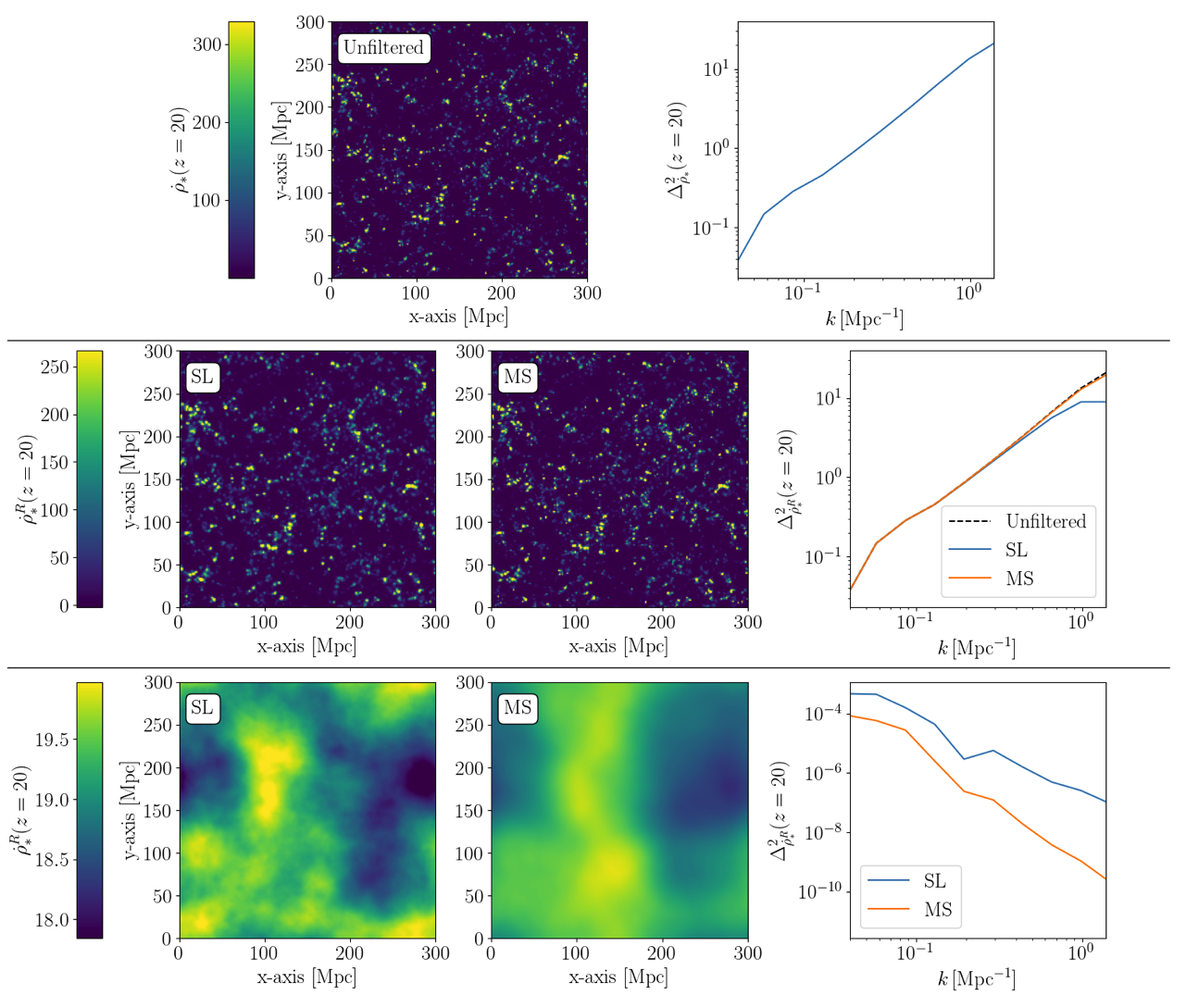}
\caption{Comparison between SL filtering and MS filtering. \textbf{Top row} shows (from left to right) the front side of the unfiltered SFRD coeval box at $z=20$ and the power spectrum of the box. \textbf{Middle row} shows (from left to right) the filtered SFRD box with the SL (MS) window function, Eq.~\eqref{eq: 11} and Eq.~\eqref{eq: 13} (Eq.~\ref{eq: 32}), and a comparison of the power spectra of the two filtered boxes, for $R_\mathrm{i}=0.93\,\mathrm{Mpc}$, $R_\mathrm{o}=1.09\,\mathrm{Mpc}$. \textbf{Bottom row} is similar to middle row, but for filter radii $R_\mathrm{i}=195\,\mathrm{Mpc}$, $R_\mathrm{o}=228\,\mathrm{Mpc}$. All SFRD coeval boxes are measured in units of $M_\odot\,\mathrm{Mpc^{-3}\,Myr^{-1}}$.}
\label{fig: 9}
\end{centering}
\end{figure*}

\begin{figure*}
\begin{centering}
\includegraphics[width=\textwidth]{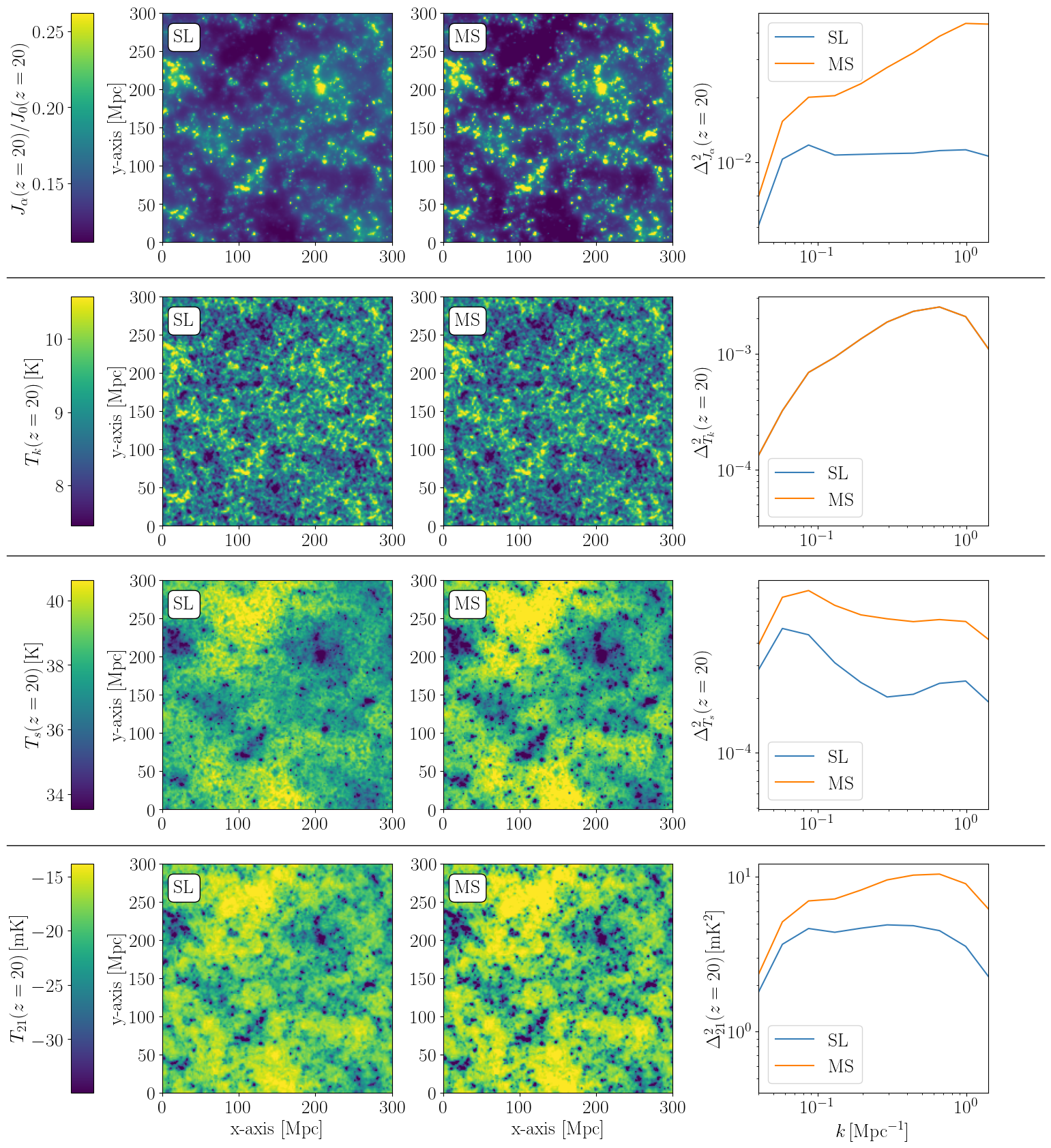}
\caption{Comparison between SL filtering and MS filtering on the relevant physical fields at $z=20$.  \textbf{Top row} shows (from left to right) the front side of the coeval box of $J_\alpha\left(z\right)/J_0\left(z\right)$ at $z=20$ when the SL filter is applied on the SFRD field (see Fig.~\ref{fig: 8}), followed by a similar plot but when the MS filtered is considered instead,  and a comparison of the power spectra of the two boxes. The redshift-dependent normalization for $J_\alpha$ is $J_0\left(z\right)=5.6\times10^{-12}\left(1+z\right)\,\mathrm{cm^{-2}s^{-1}Hz^{-1}sr^{-1}}$, this normalization was chosen since $J_\alpha\left(z\right)/J_0\left(z\right)$ is the Ly$\alpha$ coupling coefficient $x_\alpha$ in Eq.~\eqref{eq: 2}, up to a correction of order unity. \textbf{Lower rows} are similar to the top row, but display instead, in a descending order, the coeval boxes and power spectra for the gas kinetic temperature $T_k$, the spin temperature $T_s$ and the brightness temperature $T_{21}$.}
\label{fig: 10}
\end{centering}
\end{figure*}

\begin{figure*}
\begin{centering}
\includegraphics[width=\textwidth]{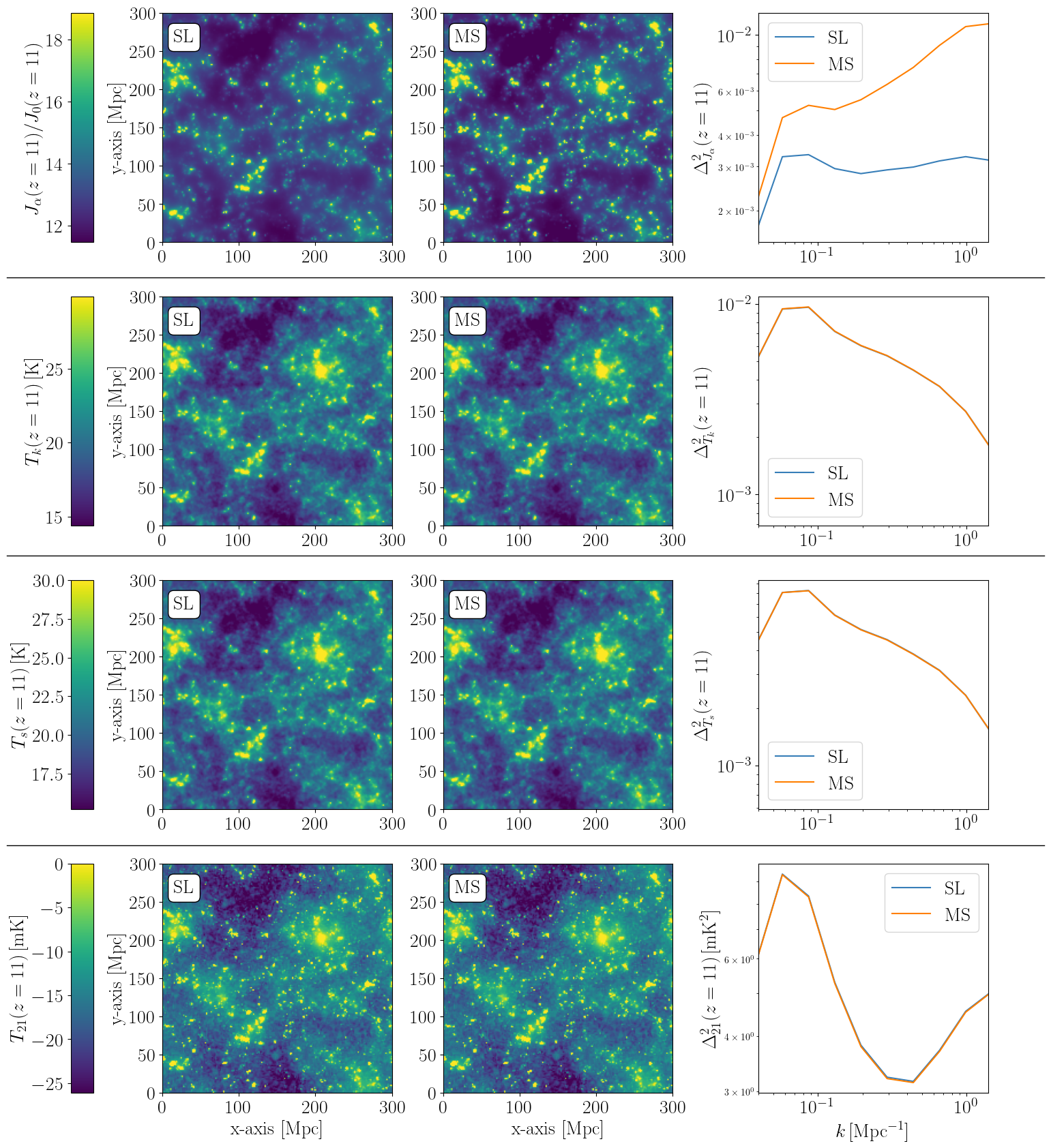}
\caption{Similarly as Fig.~\ref{fig: 10}, but all panels here correspond to $z=11$, where $J_\alpha/J_0\sim\mathcal O\left(10\right)$, thus $T_s\approx T_k$ and $T_{21}$ is insensitive to the fluctuation pattern in $J_\alpha$. }
\label{fig: 11}
\end{centering}
\end{figure*}

\begin{figure*}
\begin{centering}
\includegraphics[width=0.9\textwidth]{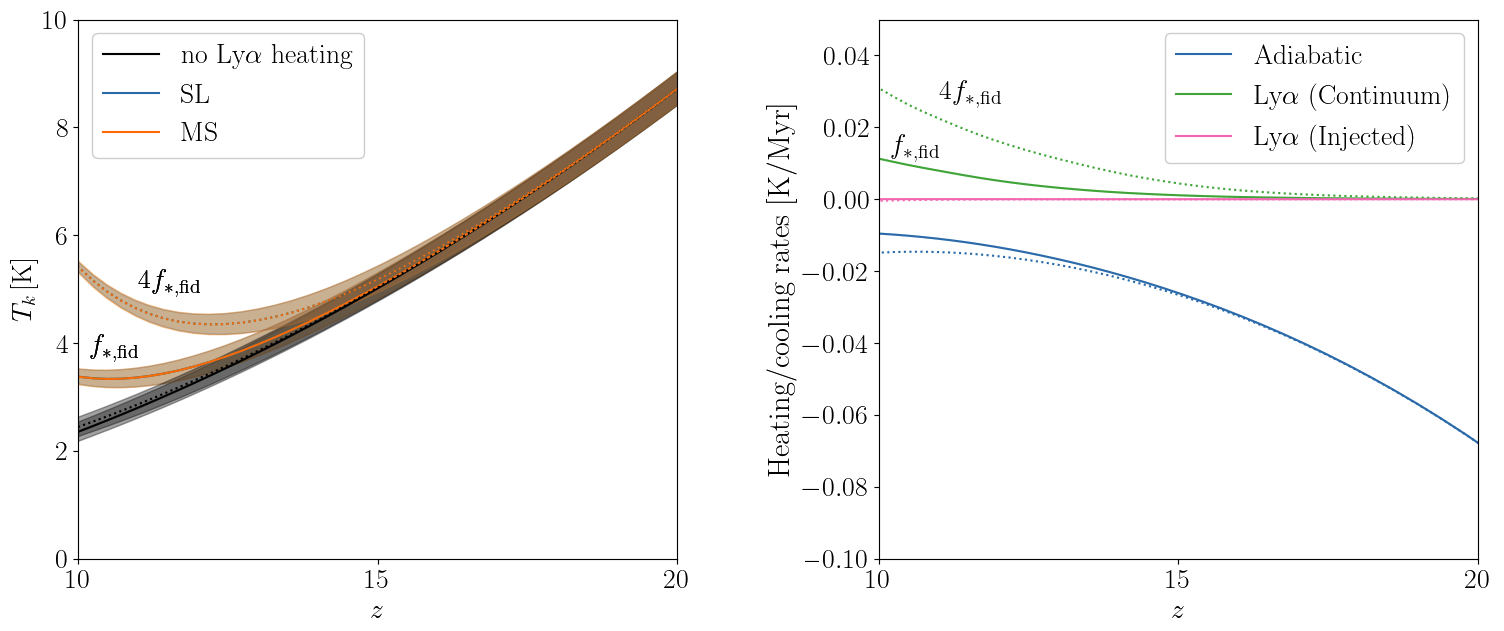}
\caption{\textbf{Left panel}: evolution of the mean $T_k$ as a function of redshift, for two different star formation efficiencies (in our fiducial model, $f_{*,\mathrm{fid}}^\mathrm{(II)}=1\%$, $f_{*,\mathrm{fid}}^\mathrm{(III)}=0.03\%$), while assuming no X-ray heating. The black curve corresponds to a scenario with no Ly$\alpha$ heating, while the blue (orange) curves account for Ly$\alpha$ heating, with SL (MS) filtering. Shaded regions around the mean value have width of the RMS of the $T_k$ field. In this panel, the blue SL curve overlaps with the orange MS curve. \textbf{Right panel}: Mean value of heating and cooling rates as a function of redshift, while Ly$\alpha$ MS is turned on. We present the adiabatic cooling rate with \emph{blue}, heating rate by continuum Ly$\alpha$ photons with \emph{green}, and cooling rate by injected Ly$\alpha$ photons with \emph{pink}. Compton and CMB heating rates~\citep{Meiksin:2021cuh} are not shown in this panel as they are negligible compared to the other rates. In both panels, \emph{solid (dotted)} lines correspond to setting $f_*=f_{*,\mathrm{fid}}$ ($f_*=4f_{*,\mathrm{fid}}$), for both popII and popIII stars.}
\label{fig: 12}
\end{centering}
\end{figure*}

Power spectra in our analysis are computed with the {\tt tuesday}\footnote{\href{https://github.com/21cmfast/tuesday}{github.com/21cmfast/tuesday}} package. For a given field $X$, we consider the dimensionless power spectrum,
\begin{equation}\label{eq: 33}
\Delta_X^2=\frac{k^3P_X\left(k\right)}{2\pi^2},
\end{equation}
where $P_X\left(k\right)$ is the angle-averaged Fourier transform of the two-point correlation function, $\langle\delta_X\left(\mathbf x\right)\delta_X\left(\mathbf x\right)\rangle$, while $\delta_X$ is the local contrast in the field $X$, $\delta_X\left(\mathbf x\right)\equiv X\left(\mathbf x\right)/\bar X-1$. For the brightness temperature, to align with the quantity that is often referred in the literature, we consider the dimensionfull 21-cm power spectrum $\Delta_{21}^2$, which is achieved by multiplying the RHS of Eq.~\eqref{eq: 33} with the square of the mean of the coeval box.

To understand how the Ly$\alpha$ MS affects the fields in the {\tt 21cmFAST} simulation, we explore first how the MS window function of Eq.~\eqref{eq: 32} (in contrast with the SL window function of Eq.~\ref{eq: 13}) affects the filtered SFRD field $\dot\rho^R_*\left(\mathrm{x},z\right)$, as shown in Fig.~\ref{fig: 9}. Firstly, notice that the unfiltered SFRD field $\dot\rho_*\left(\mathrm{x},z\right)$ is spatially intermittent, as star formation activity is significantly increased in rare massive halos, resulting in more power in fluctuations on small scales (high $k$) compared to fluctuations on large scales (low $k$). We can then see in the second row of Fig.~\ref{fig: 9} how the SL and MS window functions shape the filtered SFRD field on small distances (small $kR_\mathrm{i}$, $kR_\mathrm{o}$). On very large scales the power spectra of the filtered SFRD field remains almost the same as the power spectrum of the unfiltered SFRD field because the window function approaches unity when $k$ is very small. However, for sufficiently large $k$, the differences in the window functions yield differences in the power spectra of the filtered SFRD fields. As can be seen in Fig.~\ref{fig: 8}, the MS window function preserves more power at small $k$ values compared to the SL window function, thereby resulting in the power spectrum of the filtered SFRD with the SL window function to be suppressed, compared to the case when the MS window function is applied instead. Consequently, the filtered SFRD field appears to be smoother in the SL scenario. When considering the differences between the SL and MS scenarios on large distances, the conclusions are the opposite; Fig.~\ref{fig: 8} shows that the on the large $kR$ limit, the MS window function decays to zero more quickly compared to the SL window function. As a result, in the large distance limit, the power spectrum of the filtered SFRD in the MS scenario is suppressed by several orders of magnitude compared to the SL scenario, which leads to a much smoother filtered SFRD in the MS scenario.

Ultimately, when the Ly$\alpha$ flux $J_\alpha$ is computed, the contributions of the small filter radii dominates the contributions from the large filter radii. This is simply because the emissivity (or SFRD) field increases in magnitude at early stages of cosmic dawn and therefore filtered emissivities that are evaluated at short retarded times (correspond to small filter radii) contribute more to $J_\alpha$, c.f.~Eq.~\eqref{eq: 3}. Following the above discussion, the enhancement in power in the filtered SFRD field due to multiple scattering (on small filter radii) is therefore also evident in the power spectrum of $J_\alpha$, though on all scales, as shown in the top row of Fig.~\ref{fig: 10}. This leads to a larger contrast in the $J_\alpha$ image in the MS scenario compared to the SL scenario; regions with more (less) intensity become more (less) intense due to the effect of Ly$\alpha$ multiple-scattering.

We also show in Fig.~\ref{fig: 10} how the differences in $J_\alpha$ due to relaxation of the SL approximation are propagated to other cosmological fields at the early stages of cosmic dawn. At $z=20$, X-ray and Ly$\alpha$ heating have yet to become efficient, and so the fluctuations in the gas kinetic temperature $T_k$ at this stage follow the underlying density fluctuations due to adiabatic cooling/heating; over- (under-) dense regions imply hotter (colder) regions. Yet, the magnitude of these temperature fluctuations, at $z=20$ is subdominant to the fluctuations in $J_\alpha$. Therefore, the fluctuations pattern of the spin temperature $T_s$ at $z=20$ greatly follows the fluctuations pattern of $J_\alpha$; based on Fig.~\ref{fig: 10}, it could be argued that the fluctuations pattern in $T_s$ is (to first oder) the negative image of the fluctuations pattern in $J_\alpha$. This is physically reasonable as the spin temperature is coupled more (less) strongly to $T_k$ in regions with more (less) Ly$\alpha$ radiation, thereby pushing more (less) $T_s$ to lower values. The strong anti-correlations between $T_s$ and $J_\alpha$ cause the higher contrast feature in the latter, due to Ly$\alpha$ MS, to appear in the former as well. The same feature can also be seen when comparing the coeval boxes of the brightness temperature with and without the effect of Ly$\alpha$ MS. Relaxing the SL approximation appears to increase the 21-cm power spectrum by $\mathcal O\left(50\%\right)$, at all scales, at high redshifts such as $z=20$, when $x_\alpha\sim J_\alpha/J_0$ is less than unity.

How does Ly$\alpha$ MS affect the brightness temperature at lower redshifts? We show in Fig.~\ref{fig: 11} a similar plot of the fields with and without the SL approximation, but at $z=11$. Here, unlike Fig.~\ref{fig: 10}, $J_\alpha/J_0$ is of order 10. This means that $T_s\approx T_k$. 
Since for our fiducial astrophysics, Ly$\alpha$ heating is negligible compared to X-ray heating, the fluctuations in $T_k$ are dominated by X-ray heating, with almost no dependence on the fluctuations in $J_\alpha$. Therefore, Ly$\alpha$ MS has no effect on $T_k$, nor on $T_s$, at lower redshifts. As a consequence, the brightness temperature field and the 21-cm power spectrum at $z=11$ are not modified as well by Ly$\alpha$ MS. At even lower redshifts ($z\lesssim10$), the 21-cm power spectrum is still insensitive to the effect of Ly$\alpha$ MS, but due to another reason. At sufficiently low redshifts, $T_s\gg T_\gamma$ as $T_s\approx T_k$ and the gas is further heated by X-rays above the CMB temperature. This is the saturation limit in which $T_{21}\propto1-T_\gamma/T_s$ becomes independent on $T_s$. Since the brightness temperature does not depend on $T_s$ in this redshift regime, it does not depend on $J_\alpha$ neither, nor on the modifications of Ly$\alpha$ MS to $J_\alpha$.

\subsection{Ly$\alpha$ heating with Ly$\alpha$ multiple scattering}\label{subsec: Lya heating and Lya multiple scattering}

The interaction between Ly$\alpha$ photons and the hydrogen atoms in the IGM does not only couple between $T_s$ and the Ly$\alpha$ color temperature $T_\alpha$ via the Wouthuysen-Field effect~\citep{1952AJ.....57R..31W, 1958PIRE...46..240F}, but also couples between $T_\alpha$ and $T_k$ through energy exchange between the two heat reservoirs (though spin exchange in hydrogen atoms yields order of Kelvin differences between the two temperatures). The overall energy exchange rate has contributions from both continuum and injected photons, where the net effect of the former (latter) is to lose (gain) energy to (from) the IGM. In equilibrium, a detailed analysis can show that the heating effect of the continuum photons is more dominant than the cooling effect of the injected photons, thereby the IGM is heated in this process~\citep{Chen:2003gc, Mittal:2020kjs, Reis:2021nqf}. This heating mechanism is known as Ly$\alpha$ heating.

The two heating and cooling rates described above are proportional to the number density of continuum and injected photons, respectively. Therefore, the Ly$\alpha$ heating rate can be written as being proportional to $\Delta E_\mathrm{c}J_\mathrm{c}+\Delta E_\mathrm{i}J_\mathrm{i}$, where $J_\mathrm{c}$ and $J_\mathrm{i}$ are the continuum and injected photons ($J_\alpha=J_\mathrm{c}+J_\mathrm{i}$), and $\Delta E_\mathrm{c}$, $\Delta E_\mathrm{i}$ are functions of $T_k$, $T_s$ and the Gunn-Peterson optical depth. Thus, roughly speaking, Ly$\alpha$ heating rate is proportional to $J_\alpha$, meaning that it becomes efficient at low redshifts, as $J_\alpha$ increases. Since we have seen that MS enhances the contrast in the spatial fluctuations in $J_\alpha$, it should therefore cause a similar effect when Ly$\alpha$ heating is considered; over- (under-) dense regions are more (less) heated due to Ly$\alpha$ MS. Yet, this combined effect can become relevant only if Ly$\alpha$ heating is efficient enough to overcome any other heating and cooling mechanisms. In our fiducial model, $\log_{10}\left(L_X/\mathrm{SFR}\right)=40.5$, hence X-ray heating is much more dominant than Ly$\alpha$ heating, and the effect of Ly$\alpha$ heating on $T_k$ is barely visible in Fig~\ref{fig: 11}.

To see more clearly the combined effect of Ly$\alpha$ MS and Ly$\alpha$ heating, in the left panel of Fig.~\ref{fig: 12} we show the evolution of the global $T_k$ while turning off the effect of X-ray heating. Interestingly, even for this extreme scenario which has been ruled out by HERA~\citep{HERA:2021noe}, we see that Ly$\alpha$ heating is not very efficient in heating the gas, and that Ly$\alpha$ MS has very little effect in changing the global evolution or RMS of $T_k$. This conclusion remains the same even if we increase the normalization of the star formation efficiency $f_*$ (in both popII and popIII stars in the simulation) to be four times greater than in our fiducial model. These findings are also supported by the right panel of Fig.~\ref{fig: 12}, where the various heating and cooling rates are displayed.

If on the one hand $f_*$ is low, Ly$\alpha$ heating is very inefficient and the effect of Ly$\alpha$ MS is less visible on $T_k$. On the other hand, if $f_*$ is high, X-ray heating increases $T_k$ (and $T_s$, which is strongly more coupled to $T_k$ in that scenario) above the CMB temperature such that the saturation limit holds and $T_{21}$ does not depend on $T_s$ and $J_\alpha$. Either way, the effect that Ly$\alpha$ MS has on the brightness temperature at low redshifts is negligible, even for inefficient X-ray heating. This conclusion is in tension with the results shown in RFB21\footnote{Despite such models being ruled out by UV luminosity functions (e.g. \cite{Park:2019aul}), we have also made simulations assuming a mass-independent star-formation efficiency as is done in RFB21.  Even so, we were unable to reproduce their qualitative trends.}, as their global $T_k$ can reach $100\,\mathrm{K}$ at $z=6$ with only Ly$\alpha$ heating, while disabling X-ray heating completely. A clear resolution to this discrepancy will become available only after a thorough comparison between the two codes.


\section{Conclusions}\label{sec: Conclusions}

In this work, we presented a semi-analytic formalism to account for Ly$\alpha$ MS in 21-cm calculations. Our formalism is founded on extending the SL window function, Eq.~\eqref{eq: 13}, to any form of a normalized radial distribution that describes the probability that a photon emitted at redshift $z_\mathrm{em}$ and absorbed at redshift $z_\mathrm{abs}$ will traverse an effective comoving distance $r$. To find the radial distributions, we developed a publicly available package, {\tt SP$\alpha$RTA}, that simulates the trajectories of Ly$\alpha$ photons that are absorbed by the IGM.

We found that for any $z_\mathrm{abs}$, the radial distributions can be modeled well with analytical beta distributions that are governed by a single parameter, $x_\mathrm{em}\equiv R_\mathrm{SL}\left(z_\mathrm{abs},z_\mathrm{em}\right)/R_*\left(z_\mathrm{abs}\right)$, where $R_\mathrm{SL}\left(z_\mathrm{abs},z_\mathrm{em}\right)$ is the comoving distance between $z_\mathrm{abs}$ and $z_\mathrm{em}$ (Eq.~\ref{eq: 7}) and $R_*\left(z_\mathrm{abs}\right)$ is the comoving diffusion scale of Ly$\alpha$ photons (Eq.~\ref{eq: 24}). We showed that in the large $x_\mathrm{em}$ limit, our MS window function, Eq.~\eqref{eq: 32}, approaches the SL window function, as can be predicted from physical reasoning (this can be nicely seen in Fig.~\ref{fig: 8}). We have not varied the cosmological parameters in our analysis as this is beyond the scope of this paper. Nevertheless, because of the explicit appearance of the cosmological parameters in Eq.~\eqref{eq: 24}, and the analytical insights we have gathered on the Ly$\alpha$ MS problem in this work, we can cautiously predict that our results hold for other cosmologies as well, possibly even to extensions beyond $\Lambda$CDM. We leave however the exploration of how different cosmologies affect our formalism to future work.

By replacing the SL window function in {\tt 21cmFASTv4} with the MS window function, we could study the effect that Ly$\alpha$ MS has on the 21-cm signal. We found that the MS window function tends to increase the spatial contrast in the Ly$\alpha$ flux $J_\alpha$. This feature however only propagates to the brightness temperature at high redshifts, before $J_\alpha$ has increased to a sufficiently large value and the spin temperature follows the gas kinetic temperature. In addition, we have also explored the combined effect of Ly$\alpha$ heating and Ly$\alpha$ MS by removing X-ray heating from the simulation. Our results highlight that Ly$\alpha$ heating is still not very efficient in this configuration, and that Ly$\alpha$ MS does not affect this conclusion.

Since Ly$\alpha$ MS is only important at high redshifts, its inclusion in the analysis of data from current 21-cm interferometers is less relevant, as they are increasingly more sensitive towards lower redshifts. However, next generation 21-cm observatories, such as SKA, are expected to reach a sensitivity where the effect of Ly$\alpha$ MS can no longer be ignored. This work thus presents the necessary tools for more accurately analyzing the 21-cm signal at the onset of cosmic dawn, and thus to unearth the Lyman-alpha signature of the very first stars in the Universe.


\begin{acknowledgements}
We thank Steven G. Murray, Benoit Semelin, Shikhar Mittal, Aaron Smith and Dominic Agius for useful discussions. We also thank the anonymous referee for providing suggestions that improved the quality of this paper. We gratefully acknowledge computational resources of the HPC center at SNS. JF acknowledges support by SNS. JBM acknowledges support from NSF Grants AST-2307354 and AST-2408637, and the CosmicAI institute AST-2421782. AM acknowledges support from the Italian Ministry of Universities and Research (MUR) through the PRIN project "Optimal inference from radio images of the epoch of reionization", and the PNRR project "Centro Nazionale di Ricerca in High Performance Computing, Big Data e Quantum Computing".
\end{acknowledgements}

\appendix


\section{Derivation of the finite shell window function}\label{sec: Derivation of the finite shell window function}

In this section we write our derivation for the finite shell window function. We begin by plugging the cumulative radial distribution, Eq.~\eqref{eq: 10}, in the expression for the window function, Eq.~\eqref{eq: 8}. This gives (we omit the temporal dependence of the window function to reduce clutter)
\begin{flalign}\label{eq: A1}
&\nonumber\tilde W\left(kR_\mathrm{o},kR_\mathrm{i}\right)=\frac{\int_{0}^{\infty}dr\,r^{2}\frac{\sin kr}{kr}\int_{R_\mathrm{i}}^{R_\mathrm{o}}dr'f_{r'}\left(r\right)}{\int_{0}^{\infty}dr\,r^{2}\int_{R_\mathrm{i}}^{R_\mathrm{o}}dr'f_{r'}\left(r\right)}&
\\&\hspace{21mm}=\frac{G\left(k,R_\mathrm{o}\right)-G\left(k,R_\mathrm{i}\right)}{G\left(0,R_\mathrm{o}\right)-G\left(0,R_\mathrm{i}\right)},&
\end{flalign}
where we defined
\begin{equation}\label{eq: A2}
G\left(k,R\right)\equiv \int_{0}^{\infty}dr\,r^{2}\frac{\sin kr}{kr}\int_{0}^{R}dr'f_{r'}\left(r\right).
\end{equation}
There are two ways to proceed from here. We can either perform the $r'$ integral first and then the $r$ integral, or vice versa. We begin with the first approach. Here, we shall plug our MS beta distribution of Eq.~\eqref{eq: 26} (with the proper normalization)
\begin{flalign}\label{eq: A3}
&\nonumber G_\mathrm{MS}\left(k,R\right)=&
\\&\int_{0}^{\infty}dr\,r^{2}\frac{\sin kr}{kr}\int_{0}^{R}dr'\frac{r^{\alpha-1}\left(r'-r\right)^{\beta-1}\mathcal H\left(r'-r\right)}{\mathrm B\left(\alpha,\beta\right)r'^{\alpha+\beta-1}},&
\end{flalign}
where $\mathcal H\left(\cdot\right)$ is the Heaviside function. It is straightforward to show that the $r'$ integral in Eq.~\eqref{eq: A3} can be expressed as
\begin{flalign}\label{eq: A4}
&\nonumber\int_{0}^{R}dr'\frac{r^{\alpha-1}\left(r'-r\right)^{\beta-1}\mathcal H\left(r'-r\right)}{\mathrm B\left(\alpha,\beta\right)r'^{\alpha+\beta-1}}&
\\&\nonumber\hspace{10mm}=\nonumber\mathcal{H}\left(R-r\right)\int_{r/R}^{1}\frac{d\xi}{\xi}\frac{\xi^{\alpha-1}\left(1-\xi\right)^{\beta-1}}{B\left(\alpha,\beta\right)}&
\\&\hspace{10mm}\approx\mathcal{H}\left(R-r\right)\frac{\alpha-1}{\alpha+\beta-1}\left[1-I_{r/R}\left(\alpha-1,\beta\right)\right],&
\end{flalign}
where $\xi\equiv r/r'$ and $I_x\left(\alpha,\beta\right)$ is the regularized incomplete beta function (it is the cumulative distribution function of the beta distribution). Note the approximation made in the last line of Eq.~\eqref{eq: A3}; there is an implicit assumption that $\alpha$ and $\beta$ are constants. This is however not precise as $\alpha$ and $\beta$ depend on $x_\mathrm{em}$ and the latter depends on the maximum SL distance (given by $r'$ for the cumulative radial distribution). Yet, we shall continue with the derivation and address that nuance later in this appendix. Plugging Eq.~\eqref{eq: A4} in Eq.~\eqref{eq: A3} then yields
\begin{flalign}\label{eq: A5}
&\nonumber G_\mathrm{MS}\left(k,R\right)=&
\\&\hspace{10mm}\frac{\alpha-1}{\alpha+\beta-1}R^{3}\int_{0}^{1}dy\,y^{2}\frac{\sin kRy}{kRy}\left[1-I_{y}\left(\alpha-1,\beta\right)\right],&
\end{flalign}
which means that the MS window function can be expressed as
\begin{equation}\label{eq: A6}
\tilde W_\mathrm{MS}\left(kR_\mathrm{o},kR_\mathrm{i}\right)=\frac{R_{\mathrm{o}}^{3}\,M_\mathrm{MS}\left(kR_{\mathrm{o}}\right)-R_{\mathrm{i}}^{3}\,M_\mathrm{MS}\left(kR_{\mathrm{i}}\right)}{R_{\mathrm{o}}^{3}-R_{i}^{3}},
\end{equation}
where 
\begin{flalign}\label{eq: A7}
&\nonumber M_\mathrm{MS}\left(kR\right)\equiv\frac{G_\mathrm{MS}\left(k,R\right)}{G_\mathrm{MS}\left(0,R\right)}&
\\&\hspace{15mm}=\frac{\int_{0}^{1}dy\,y^{2}\frac{\sin kRy}{kRy}\left[1-I_{y}\left(\alpha-1,\beta\right)\right]}{\int_{0}^{1}dy\,y^{2}\left[1-I_{y}\left(\alpha-1,\beta\right)\right]}.&
\end{flalign}
One can verify that indeed $\lim_{\alpha/\beta\to\infty} M_\mathrm{MS}\left(kR\right)=M_\mathrm{SL}\left(kR\right)$, where $M_\mathrm{SL}\left(kR\right)$ is given by Eq.~\eqref{eq: 13}, as $\lim_{\alpha/\beta\to\infty}I_{y}\left(\alpha,\beta\right)=0$ for every $y\in\left[0,1\right]$. This confirms our understanding that $\alpha\gg\beta$ corresponds to the SL scenario.

Since Eq.~\eqref{eq: A7} is not an analytical function, we take now the second approach of performing first the $r$ integral in Eq.~\eqref{eq: A3} and then the $r'$ integral. This gives
\begin{flalign}\label{eq: A8}
&\nonumber G\left(k,R_\mathrm{o}\right)-G\left(k,R_\mathrm{i}\right)=\int_{R_\mathrm{i}}^{R_\mathrm{o}}dr'\int_{0}^{\infty}dr\,r^{2}\frac{\sin kr}{kr}f_{r'}\left(r\right)&
\\&\nonumber\hspace{10mm}=\int_{R_\mathrm{i}}^{R_\mathrm{o}}dr'\,W\left(kr'\right)\int_{0}^{r'}dr\,r^{2}f_{r'}\left(r\right),&
\\&\hspace{10mm}=\int_{R_\mathrm{i}}^{R_\mathrm{o}}dr'\,r'^2\,W\left(kr'\right)\int_{0}^{1}d\xi \xi^2 r'f_{r'}\left(\xi r'\right),&
\end{flalign}
where in the second equality we used the definition of the window function (Eq.~\ref{eq: 8}) and switched again variables, $\xi\equiv r/r'$. In order to proceed with the derivation, we now approximate $r'f_{r'}\left(\xi r'\right)\approx f_1\left(\xi\right)$. This approximation works as long as the shape of $f_{r'}\left(\xi r'\right)$ doesn't change considerably for $R_\mathrm{i}\leq r'\leq R_\mathrm{o}$. In other words, we assume that in the MS scenario $\alpha$ and $\beta$ do not change much for $R_\mathrm{i}\leq R_\mathrm{SL}\leq R_\mathrm{o}$. This condition is met in our {\tt 21cmFAST} simulations as the radii of the shells satisify $R_\mathrm{o}-R_\mathrm{i}\simeq0.17 \,R_\mathrm{i}$. With this approximation, the $\xi$ integral in Eq.~\eqref{eq: A8} becomes a constant that is canceled out in Eq.~\eqref{eq: A1}. Upon changing variables, $x\equiv kr'$, we arrive at
\begin{flalign}\label{eq: A9}
&\nonumber\tilde W\left(kR_\mathrm{o},kR_\mathrm{i}\right)=\frac{k^{-3}\int_{kR_\mathrm{i}}^{kR_\mathrm{o}}dx\,x^2\,W\left(x\right)}{\int_{R_\mathrm{i}}^{R_\mathrm{o}}dr'\,r'^2\,W\left(0\right)}&
\\&\hspace{20mm}=\frac{R_{\mathrm{o}}^{3}\,M\left(kR_{\mathrm{o}}\right)-R_{\mathrm{i}}^{3}\,M\left(kR_{\mathrm{i}}\right)}{R_{\mathrm{o}}^{3}-R_{i}^{3}},&
\end{flalign}
where we used the fact that $W\left(0\right)\equiv1$ and defined
\begin{equation}\label{eq: A10}
M\left(kR\right)\equiv\frac{3}{k^3R^3}\int_0^{kR}dxx^{2}W\left(x\right).
\end{equation}

As was discussed at Sec.~\ref{subsec: Analytical fit}, the resulting window function for the MS scenario is
\begin{flalign}\label{eq: A11}
&\nonumber M_\mathrm{MS}\left(kR\right)=&
\\&{_{2}F_{3}}\left(\frac{2+\alpha}{2},\frac{3+\alpha}{2};\frac{5}{2},\frac{2+\alpha+\beta}{2},\frac{3+\alpha+\beta}{2};-\frac{1}{4}k^2R^2\right).&
\end{flalign}
Interestingly, Eqs.~\eqref{eq: A7} and \eqref{eq: A11} form a non-trivial mathematical identity. However, the hypergeometric function in Eq.~\eqref{eq: A11} is an analytical function, making its implementation in {\tt 21cmFAST} much less computationally expensive. 


\section{Conditional Gaussian random variables}\label{sec: Conditional Gaussian random variables}

In linear perturbation theory with Gaussian initial conditions, all cosmological fields have Gaussian 
probability density functions (PDFs), namely if we consider the distribution of some field $A$ at point $\mathbf x$ and redshift $z$, then $A\left(\mathbf{x},z\right)$ distributes normally with mean $\mu_{A}\left(z\right)\equiv0$ and variance $\sigma_{A}^{2}\left(z\right)$ that can be computed analytically (see some examples at Appendix \ref{sec: Cosmological correlation functions}). Consider now a group of fields that are evaluated at different points in spacetime,
$\mathbf{Y}=\Big[A\left(\mathbf{x}_{1},z_{1}\right),\,B\left(\mathbf{x}_{1},z_{1}\right),\,C\left(\mathbf{x}_{1},z_{1}\right),\cdots,A\left(\mathbf{x}_{M1},z_{M1}\right),\\B\left(\mathbf{x}_{M2},z_{M2}\right),C\left(\mathbf{x}_{M3},z_{M3}\right),\,\cdots\Big]^{\mathrm{T}}$. The joint distribution of all the elements of the vector $\mathbf{Y}$ is a multivariate Gaussian,
\begin{flalign}\label{eq: B1}
&\nonumber f_{\mathbf{Y}}\left(\mathbf{Y}=\mathbf{y}\right)=\frac{1}{\sqrt{\left(2\pi\right)^{N}\det\boldsymbol\Sigma}}&
\\&\hspace{20mm}\times\exp\left(-\frac{1}{2}\left(\mathbf{y}-\mathbf{\boldsymbol{\mu}}\right)^{\mathrm{T}}\boldsymbol{\Sigma}^{-1}\left(\mathbf{y}-\mathbf{\boldsymbol{\mu}}\right)\right),&
\end{flalign}
where $\boldsymbol{\mu}\equiv\boldsymbol0$ is a vector containing the zero mean values of the random vector $\mathbf{Y}$, $N$ is the dimension of $\mathbf{Y}$, and $\boldsymbol{\Sigma}$ is the covariance matrix. It is well known that in order for $f_{\mathbf{Y}}$ to be properly normalized, the covariance matrix $\boldsymbol{\Sigma}$ has to be positive-definite, namely $\mathbf{z}^{\mathrm{T}}\mathbf{\Sigma}\mathbf{z}>0$ for any $\mathbf{z}\in\mathbb{R}^{N}$.

Let us suppose that the random vector $\mathbf{Y}$ can be decomposed to two sub-vectors $\mathbf{Y}=\left[\mathbf{Y}_{1},\,\mathbf{Y}_{2}\right]^{\mathrm{T}}$, with mean $\boldsymbol{\mu}=\left[\mathbf{\boldsymbol{\mu}}_{1},\,\boldsymbol{\mu}_{2}\right]$ and covariance matrix $\boldsymbol{\Sigma}=\left[\begin{matrix}
\boldsymbol{\Sigma}_{11} & \boldsymbol{\Sigma}_{12}\\
\boldsymbol{\Sigma}_{21} & \boldsymbol{\Sigma}_{22}
\end{matrix}\right]$. Then the conditional PDF for $\mathbf{Y}_{1}$ given $\mathbf{Y}_{2}=\mathbf{y}_{2}$ is also a multivariate Gaussian with a conditional mean
\begin{equation}\label{eq: B2}
\boldsymbol{\mu}_\mathrm{cond}=\boldsymbol{\mu}_{1}+\boldsymbol{\Sigma}_{12}\boldsymbol{\Sigma}_{22}^{-1}\left(\mathbf{y}_{2}-\boldsymbol{\mu}_{2}\right)=\boldsymbol{\Sigma}_{12}\boldsymbol{\Sigma}_{22}^{-1}\mathbf{y}_{2},
\end{equation}
and a conditional covariance matrix
\begin{equation}\label{eq: B3}
\boldsymbol{\Sigma}_\mathrm{cond}=\boldsymbol{\Sigma}_{11}-\boldsymbol{\Sigma}_{12}\boldsymbol{\Sigma}_{22}^{-1}\boldsymbol{\Sigma}_{21}.
\end{equation}

As a concrete example, let us suppose that $\mathbf{Y}$ contains only two Gaussian random variables, $\mathbf{Y}=\left[A,\,B\right]^{\mathrm{T}}$. In the context of this work, $A$ and $B$ can be the realizations of one of the components of the peculiar velocity field in neighboring samples (thus $\mu_{A}=\mu_{B}=0$). The $2\times2$ joint covariance matrix in this case is
\begin{equation}\label{eq: B4}
\mathbf{\Sigma}=\left[\begin{matrix}
\sigma_{A}^{2} & \rho_{AB}\sigma_{A}\sigma_{B}\\
\rho_{AB}\sigma_{A}\sigma_{B} & \sigma_{B}^{2}
\end{matrix}\right]
\end{equation}
where $\rho_{AB}$ is the Pearson correlation coefficient, formally defined as
\begin{flalign}\label{eq: B5}
&\nonumber\rho_{AB}\equiv\frac{\mathrm{cov}\left(A,B\right)}{\sigma_{A}\sigma_{B}}=\frac{\langle AB\rangle-\mu_{A}\mu_{B}}{\left[\langle A^{2}\rangle-\mu_{A}\right]^{1/2}\left[\langle B^{2}\rangle-\mu_{B}\right]^{1/2}}&
\\&\hspace{5mm}=\frac{\langle AB\rangle}{\langle A^{2}\rangle^{1/2}\langle B^{2}\rangle^{1/2}}.&
\end{flalign}
The determinant of the covariance matrix in Eq.~\eqref{eq: B4} is $\left(1-\rho_{AB}^{2}\right)\sigma_{A}^{2}\sigma_{B}^{2}$, implying that $\boldsymbol{\Sigma}$ is positive-definite if and only if $-1<\rho_{AB}<1$. In this work, since we focus on correlations at small distances, we have $\rho_{AB}\approx1$, meaning that we deal with highly correlated variables.

The conditional distribution of $A$ given $B=b$ can be shown to be
\begin{flalign}\label{eq: B6}
&\nonumber f_{A|B}\left(A=a|B=b\right)=\frac{f_{AB}\left(A=a,B=b\right)}{f_{B}\left(B=b\right)}&
\\&\nonumber\hspace{5mm}=\frac{1}{\sqrt{2\pi\sigma_{A}^{2}\left(1-\rho_{AB}^{2}\right)}}&
\\&\hspace{5mm}\times\exp\left(-\left[\frac{\left[a-\mu_{A}-\rho_{AB}\frac{\sigma_{A}}{\sigma_{B}}\left(b-\mu_{B}\right)\right]^{2}}{2\sigma_{A}^{2}\left(1-\rho_{AB}^{2}\right)}\right]\right).&
\end{flalign}
This explicitly shows that, given $B=b$, $A$ distributes as $\mathcal N\left(\mu_\mathrm{cond},\sigma_\mathrm{cond}\right)$ with
\begin{equation}\label{eq: B7}
\mu_\mathrm{cond}=\mu_{A}+\rho_{AB}\frac{\sigma_{A}}{\sigma_{B}}\left(b-\mu_{B}\right)=\rho_{AB}\frac{\sigma_{A}}{\sigma_{B}}b
\end{equation}
and
\begin{equation}\label{eq: B8}
\sigma_\mathrm{cond}=\Sigma_\mathrm{cond}^{1/2}=\sigma_{A}\left(1-\rho_{AB}^{2}\right)^{1/2}.
\end{equation}
The last two equations are a special case of Eqs.~\eqref{eq: B2}-\eqref{eq: B3}.


\section{Cosmological correlation functions}\label{sec: Cosmological correlation functions}

Suppose we have two cosmological random fields, $A$ and $B$, whose two-point cross-correlation function is of interest. Let us also suppose that the field $A$ is evaluated at point $\mathbf{x}$ and redshift $z_{1}$, while field $B$ is evaluated at redshift $z_{2}$ at a point that is located a comoving distance $r$ along the unit vector $\mathbf{n}_{||}$ with respect to $\mathbf x$. We would therefore like to compute $\left\langle A\left(\mathbf{x},z_{1}\right)B\left(\mathbf{x}+r\,\mathbf{n}_{||},z_{2}\right)\right\rangle$. Given the transfer functions $\mathcal{T}_{A}\left(k,z_1\right)\equiv A\left(\mathbf k,z_1\right)/\mathcal R\left(\mathbf k\right)$ and $\mathcal{T}_{B}\left(k,z_2\right)\equiv B\left(\mathbf k,z_2\right)/\mathcal R\left(\mathbf k\right)$, where $\mathcal R\left(\mathbf k\right)$ is the Fourier transform of the primordial curvature field, this two-point function can be computed analytically with linear perturbation theory.

\begin{flalign}\label{eq: C1}
&\nonumber\left\langle A\left(\mathbf{x},z_{1}\right)B\left(\mathbf{x}+r\,\mathbf{n}_{||},z_{2}\right)\right\rangle& 
\\&\nonumber=\Bigg\langle \int\frac{d^{3}k}{\left(2\pi\right)^{3}}A\left(\mathbf{k},z_{1}\right)\mathrm{e}^{-i\mathbf{k}\cdot\mathbf{x}}&
\\&\nonumber\hspace{20mm}\times\int\frac{d^{3}k'}{\left(2\pi\right)^{3}}B\left(\mathbf{k}',z_{2}\right)\mathrm{e}^{-i\mathbf{k'}\cdot\left(\mathbf{x}+r\mathbf{n}_{||}\right)}\Bigg\rangle&
\\&\nonumber=\iint\frac{d^{3}k\,d^{3}k'}{\left(2\pi\right)^{6}}\mathrm{e}^{-i\left(\mathbf{k}+\mathbf{k}'\right)\cdot\mathbf{x}-i\mathbf{k'}\cdot r\mathbf{n}_{||}}&
\\&\nonumber\hspace{20mm}\times\mathcal{T}_{A}\left(k,z_{1}\right)\mathcal{T}_{B}\left(k',z_{2}\right)\left\langle \mathcal{R}\left(\mathbf{k}\right)\mathcal{R}\left(\mathbf{k}'\right)\right\rangle&
\\&\nonumber=\frac{1}{4\pi}\int\frac{d^{3}k}{k^{3}}\mathrm{e}^{i\mathbf{k}\cdot r\mathbf{n}_{||}}\mathcal{T}_{A}\left(k,z_{1}\right)\mathcal{T}_{B}\left(k,z_{2}\right)\Delta_{\mathcal{R}}^{2}\left(k\right)&
\\&\nonumber=\frac{1}{4\pi}\int_{0}^{\infty}\frac{dk}{k}\Delta_{\mathcal{R}}^{2}\left(k\right)\int_{-1}^{1}d\mu&
\\&\hspace{20mm}\times\int_{0}^{2\pi}d\phi\mathcal{T}_{A}\left(k,z_{1}\right)\mathcal{T}_{B}\left(k,z_{2}\right)\mathrm{e}^{-ikr\mu},&
\end{flalign}
where we used $\left\langle \mathcal{R}\left(\mathbf{k}\right)\mathcal{R}\left(\mathbf{k}'\right)\right\rangle=\left(2\pi\right)^{3}\frac{2\pi^{2}}{k^{3}}\Delta_{\mathcal{R}}^{2}\left(k\right)\delta^{\left(3\right)}\left(\mathbf{k}+\mathbf{k}'\right)$, with $\Delta_{\mathcal{R}}^{2}\left(k\right)=A_s\left(k/k_\star\right)^{n_s-1}$ the dimensionless primordial curvature power spectrum, parameterized with amplitude $A_s$ at the pivot wavenumber $k_\star=0.05\,\mathrm{Mpc}^{-1}$ and a tilt $n_s$. In the last line of Eq.~\eqref{eq: C1} we decomposed the volume integral in Fourier space into a radial integral and two additional angular integrals and defined the dummy variable $\mu\equiv\left(\mathbf{k}\cdot\mathbf{n}_{||}\right)/k$.

\begin{table*}[t!]
\begin{tabular}{
|W{c}{0.02\textwidth}|
|W{c}{0.27\textwidth}
|W{c}{0.38\textwidth}
|W{c}{0.3\textwidth}|}
\hline
 & $\delta$ & $v^{||}$ & $v^\perp$ \\
\hline\hline
$\delta$ & $\displaystyle{\frac{\sin kr}{kr}\approx1-\frac{k^{2}r^{2}}{6}}$ & $\displaystyle{\frac{\sqrt{3}\left[\sin\left(kr\right)-kr\cos\left(kr\right)\right]}{k^{2}r^{2}}\approx\frac{kr}{\sqrt{3}}}$ & 0 \\
\hline
$v^{||}$ & $\displaystyle{\frac{\sqrt{3}\left[\sin\left(kr\right)-kr\cos\left(kr\right)\right]}{k^{2}r^{2}}\approx\frac{kr}{\sqrt{3}}}$ & $\displaystyle{\frac{3\left(k^{2}r^{2}-2\right)\sin\left(kr\right)+6kr\cos\left(kr\right)}{k^{3}r^{3}}\approx1-\frac{3k^{2}r^{2}}{10}}$ & 0 \\
\hline
$v^\perp$ & 0 & 0 & $\displaystyle{\frac{3\left[\sin\left(kr\right)-kr\cos\left(kr\right)\right]}{k^{3}r^{3}}\approx1-\frac{k^{2}r^{2}}{10}}$ \\[2mm]
\hline
\end{tabular}
\caption{Window functions $W_{A,B}\left(kr\right)$ for the computation of the two-point cross-correlation functions, Eq.~\eqref{eq: C3}. We also provide the Taylor expansion of these window functions for $kr\ll1$. It should be emphasized that the window function of two different $v^{\perp}$ components is zero.}
\label{tab: 1}
\end{table*}

\begin{figure}
\begin{centering}
\includegraphics[width=\columnwidth]{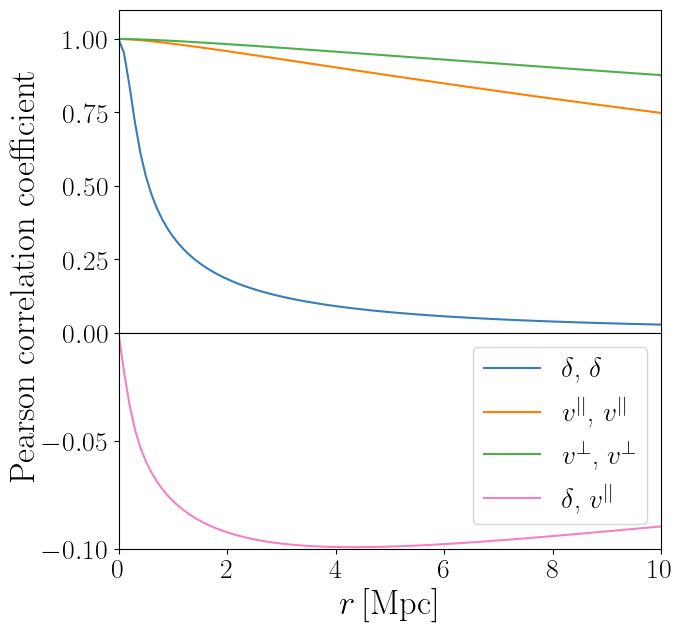}
\caption{The Pearson correlation coefficient $\rho$ as a function of distance $r$, as computed by Eq.~\eqref{eq: C5}. In all cases shown, the fields are smoothed with a top-hat filter of radius $200\,\mathrm{kpc}$, $z_1=10$, and $z_2$ is found by solving $r=R_\mathrm{SL}\left(z_1,z_2\right)$. It is worth noting that the displayed curves are insensitive to the chosen value of $z_1$, thus $\rho$ can be approximated as a function of $r$ only.} 
\label{fig: 13}
\end{centering}
\end{figure}

For the propagation of Ly$\alpha$ photons in the IGM, there are four cosmological fields of interest, the baryon over-density $\delta_{b}$ and the three components of the baryon peculiar velocity $\mathbf{v}_{b}$ (we now remove the subscript $b$ to reduce clutter). If $A=B=\delta$ , then the transfer function does not depend on either $\mu$ or $\phi$. This is not the case however when either $A$ or $B$ are the $i$'th component of $\mathbf{v}$, $v^{i}$. To see that, let us write first the transfer function for $v^{i}$.

\begin{flalign}\label{eq: C2}
&\nonumber\mathcal{T}_{v^{i}}\left(k,z\right)=\frac{\mathbf{n}^{i}\cdot\mathbf{v}\left(\mathbf{k},z\right)}{\mathcal{R}\left(\mathbf{k}\right)}=\frac{\mathbf{n}^{i}\cdot\mathbf{k}}{k}\frac{v\left(\mathbf{k},z\right)}{\mathcal{R}\left(\mathbf{k}\right)}=\frac{\mathbf{n}^{i}\cdot\mathbf{k}}{ik^{2}}\frac{\theta\left(\mathbf{k},z\right)}{\mathcal{R}\left(\mathbf{k}\right)}&
\\&\hspace{13mm}=\frac{\mathbf{n}^{i}\cdot\mathbf{k}}{k}\frac{\mathcal{T}_{\theta}\left(\mathbf{k},z\right)}{ik},&
\end{flalign}
where we used zero vorticity, $\mathbf{k}\times\mathbf{v}\left(\mathbf{k},z\right)=0$, and we used the definition for the velocity divergence in Fourier space, $\theta\left(\mathbf{k},z\right)\equiv i\mathbf{k}\cdot\mathbf{v}\left(\mathbf{k},z\right)=ikv\left(\mathbf{k},z\right)$. The result of $\left(\mathbf{n}^{i}\cdot\mathbf{k}\right)/k$ depends on which component of the velocity vector we consider. If we consider the parallel component of the velocity vector, that is the component that is parallel to $\mathbf{n}_{||}$, then $\left(\mathbf{n}^{||}\cdot\mathbf{k}\right)/k=\mu$. Otherwise, if we consider the two other perpendicular components, we get $\left(\mathbf{n}^{\perp}\cdot\mathbf{k}\right)/k=\sqrt{1-\mu^{2}}\cos\phi$ (or $\sqrt{1-\mu^{2}}\sin\phi$). From here, it is easy to see that the cross-correlation of any perpendicular component with any other field vanishes (due to the $\phi$  integral in Eq.~\ref{eq: C1}). However, it is interesting to note that the cross-correlation between $\delta$ and $v^{||}$ does not vanish, unless $r=0$. In fact, for $r>0$ this cross-correlation is negative, as can be seen in Fig.~\ref{fig: 13}. This is not surprising as over-dense regions attract more matter towards them, thereby reducing the value of the parallel velocity of particles away from them.

All in all, the various correlation functions can be computed according to
\begin{flalign}\label{eq: C3}
&\nonumber\left\langle A\left(\mathbf{x},z_{1}\right)B\left(\mathbf{x}+r\,\mathbf{n}_{||},z_{2}\right)\right\rangle&
\\&\hspace{10mm}=\int_{0}^{\infty}\frac{dk}{k}\Delta_{\mathcal{R}}^{2}\left(k\right)\tilde{\mathcal{T}}_A\left(k,z_{1}\right)\tilde{\mathcal{T}}_B\left(k,z_{2}\right)W_{A,B}\left(kr\right),&
\end{flalign}
where $\tilde{\mathcal T}_\delta\left(k,z\right)\equiv\mathcal T_\delta\left(k,z\right)$ while $\tilde{\mathcal T}_{v^i}\left(k,z\right)\equiv\mathcal T_\theta\left(k,z\right)/(\sqrt{3}k)$, and the window functions $W_{A,B}\left(kr\right)$ are listed in Table \ref{tab: 1}. From that table we can see that two perpendicular components of the peculiar velocity, $v^\perp_1$ and $v^\perp_2$, are completely uncorrelated with any other field, and thus they depend only on their own past samples. However, as discussed above, since $W_{\delta,v^{||}}\left(kr\right)\neq0$ unless $r=0$, there is a non-vanishing anti-correlation between $\delta$ and $v^{||}$.

In {\tt SP$\alpha$RTA}, we obtain the transfer functions for $\delta$ and $\theta$ from {\tt CLASS}\footnote{\href{https://github.com/lesgourg/class_public}{github.com/lesgourg/class\_public}}~\citep{Blas:2011rf}.
Since we work with a finite step-size of size $L_\mathrm{cell}$ between samples, we in fact consider a smoothed version of the cosmological fields. Therefore, in practice, we modify the transfer functions we consider in Eq.~\eqref{eq: C3} according to $\tilde{\mathcal T}_A\left(k,z\right)\to W_\mathrm{top}\left(kL_\mathrm{cell}\right)\tilde{\mathcal T}_A\left(k,z\right)$, where $W_\mathrm{top}\left(kL_\mathrm{cell}\right)$ is the Fourier transform of a top-hat filter in real space with radius $L_\mathrm{cell}$ (it has the same mathematical expression as $M_\mathrm{SL}\left(kR\right)$ in Eq.~\ref{eq: 13}).

The variance of a field $A$ at some redshift $z$ can be obtained by setting $B=A$, $z_1=z_2=z$ and $r=0$ in Eq.~\eqref{eq: C3}. Using the fact that the auto window functions were normalized such that $W_{A,A}\left(0\right)\equiv1$, we get 
\begin{equation}\label{eq: C4}
\left\langle A^2\left(\mathbf{x},z\right)\right\rangle=\int_{0}^{\infty}\frac{dk}{k}\Delta_{\mathcal{R}}^{2}\left(k\right)\tilde{\mathcal{T}}_A^2\left(k,z\right).
\end{equation}
Then, from Eq.~\eqref{eq: B5} the Pearson correlation coefficient is
\begin{flalign}\label{eq: C5}
&\nonumber\rho_{AB}=&
\\&\frac{\int_{0}^{\infty}\frac{dk}{k}\Delta_{\mathcal{R}}^{2}\left(k\right)\tilde{\mathcal{T}}_A\left(k,z_{1}\right)\tilde{\mathcal{T}}_B\left(k,z_{2}\right)W_{A,B}\left(kr\right)}{\left[\int_{0}^{\infty}\frac{dk}{k}\Delta_{\mathcal{R}}^{2}\left(k\right)\tilde{\mathcal{T}}_A^2\left(k,z_1\right)\right]^{1/2}\left[\int_{0}^{\infty}\frac{dk}{k}\Delta_{\mathcal{R}}^{2}\left(k\right)\tilde{\mathcal{T}}_B^2\left(k,z_2\right)\right]^{1/2}}.&
\end{flalign}
Note that by construction, the Pearson correlation coefficient satisfies $-1\leq\rho_{AB}\leq 1$. Specifically, for $A=B$, $z_2-z_1=\Delta z\ll z_1$, and $r=R_\mathrm{SL}\left(z_1,z_2\right)$, it is easy to see that $\rho_{A,A}\approx1-\mathcal O\left(\Delta z\right)$. For sufficiently large redshift difference (or comoving distance $r$), $\left|\rho_{A,A}\right|$ is expected in general to decay towards zero (though there could be some small exceptions, e.g. baryon acoustic oscillations).

\subsection{Consistency check}\label{subsec: Consistency check}

As was described in Sec.~\ref{sec: SPaRTA}, {\tt SP$\alpha$RTA} does not work with a grid in order to make it more computationally efficient. The idea behind this design is that there is no need to know the value of $\delta$ and $\mathbf v$ across an entire grid in order to simulate the trajectory of a single photon, but rather only at the points along its path. In theory, under the assumption that the cosmological fields begin with Gaussian initial conditions and are evolved according to linear perturbation theory, the realizations of the fields along the photon's trajectory can be drawn from the formalism presented in this and the previous appendices.

We would like to understand however what are the limitations of our analytic formalism. For that purpose, let us consider three Gaussian random variables, $A\equiv\delta\left(\mathbf{x}_{2},z_{2}\right)$, $B\equiv\delta\left(\mathbf{x}_{1},z_{1}\right)$ and $C\equiv v^{||}\left(\mathbf{x}_{1},z_{1}\right)$. Given $B$ and $C$, we would like to draw $A$. Is this possible in our formalism? To answer that question, recall that the three variables form a multivariate distribution of rank 3, and the covariance matrix of this distribution is given by
\begin{equation}\label{eq: C6}
\mathbf{\Sigma}=\left[\begin{matrix}
\sigma_{A}^{2} & \rho_{AB}\sigma_{A}\sigma_{B} & \rho_{AC}\sigma_{A}\sigma_{C}\\
\rho_{AB}\sigma_{A}\sigma_{B} & \sigma_{B}^{2} & \rho_{BC}\sigma_{B}\sigma_{C}\\
\rho_{AC}\sigma_{A}\sigma_{C} & \rho_{BC}\sigma_{B}\sigma_{C} & \sigma_{C}^{2}
\end{matrix}\right].
\end{equation}
The determinant of this matrix is
\begin{flalign}\label{eq: C7}
&\nonumber\det\boldsymbol{\Sigma}=&
\\&\hspace{10mm}\sigma_{A}^{2}\sigma_{B}^{2}\sigma_{C}^{2}\left(1-\rho_{AB}^{2}-\rho_{AC}^{2}-\rho_{BC}^{2}+2\rho_{AB}\rho_{AC}\rho_{BC}\right).&
\end{flalign}
Now, based on the above discussion, we can expect that $\rho_{BC}=0$ and $\rho_{AB}\approx1$ if $z_{2}\approx z_{1}$. This gives us a determinant of $\det\boldsymbol{\Sigma}\approx-\sigma_{A}^{2}\sigma_{B}^{2}\sigma_{C}^{2}\,\rho_{AC}^{2}<0$, namely $\boldsymbol{\Sigma}$ is not positive-definite and we cannot draw all three fields from a joint normalized multivariate Gaussian distribution.

Furthermore, let us consider the following four Gaussian random variables, $A\equiv\delta\left(\mathbf{x}_{2},z_{2}\right)$, $B\equiv v^{||}\left(\mathbf{x}_{2},z_{2}\right)$, $C\equiv \delta\left(\mathbf{x}_{1},z_{1}\right)$ and $D\equiv v^{||}\left(\mathbf{x}_{1},z_{1}\right)$. We would like to draw now $A$ and $B$ given $C$ and $D$. Based on the previous discussion, we now expect that $\rho_{AB}=\rho_{CD}=0$ while $\rho_{AC}\approx\rho_{BD}\approx1$. Using Eq.~\eqref{eq: B3}, the determinant of the conditional covariance matrix is $\det\boldsymbol{\Sigma}_\mathrm{cond}\approx-\sigma_A^2\sigma_B^2\left[\left(\rho_{AD}+\rho_{BC}\right)^2-\rho_{AD}^2\rho_{BC}^2\right]$, which is negative as long as $\rho_{AD}$ and $\rho_{BC}$ have the same sign. 

These non-positive-definite covariance matrices indicate that we are unable to \emph{consistently} draw a pair of $\delta$ and $v^{||}$ given \emph{only} the previous sample of the pair. This happens because on the one hand $\delta\left(\mathbf x_2,z_2\right)$ and $v^{||}\left(\mathbf x_2,z_2\right)$ are completely uncorrelated, so we have no information about one of them given the other. However, on the other hand, $\delta\left(\mathbf x_2,z_2\right)$ is very correlated with $\delta\left(\mathbf x_1,z_1\right)$, while $\delta\left(\mathbf x_1,z_1\right)$ is weakly anti-correlated with $v^{||}\left(\mathbf x_2,z_2\right)$, so we have somewhat information on $v^{||}\left(\mathbf x_2,z_2\right)$ given $\delta\left(\mathbf x_2,z_2\right)$. The only way to solve this contradiction is to simultaneously consider enough samples such that the information that comes from all of them is canceled out when trying to infer $\delta\left(\mathbf x_2,z_2\right)$ from $v^{||}\left(\mathbf x_2,z_2\right)$ (or vice versa). In other words, at every step in the simulation, the size of the joint covariance matrix needs to be increased until it is positive-definite. While this is technically possible, we defer the implementation of this solution to future work. 

In the meantime, in order to overcome this mathematical obstacle, in this work we have completely ignored the inhomogeneities in the density field in order to determine the radial distributions, setting effectively $\delta\equiv0$ in {\tt SP$\alpha$RTA}. Our justification to this approximation relies on the work of Ref.~\cite{Mittal:2023xih}, which have shown that the inhomogeneities in the density field have little effect on $J_\alpha$ when Ly$\alpha$ MS is taken into account.

Since in {\tt SP$\alpha$RTA} we draw the current velocity vector based on only the previous velocity vector, our realization of the velocity follows the statistics of a Gauss-Markov process. In this process, if $\rho\left(n,m\right)$ denotes the Pearson correlation coefficient between the $n$ and $m$ samples, then it satisfies the following property, $\rho\left(n,m\right)=\rho\left(n,l\right)\rho\left(l,m\right)$, where $n\leq l\leq m$. While this Markovian approximation works well on the smallest scales, where $\rho\approx1$, it tends to overestimate $\rho$ on larger scales, resulting in smaller relative velocities compared to what linear perturbation theory predicts. To remedy this bias towards lower relative velocities, while keeping the speedy analytical calculations of {\tt SP$\alpha$RTA}, we multiply in the code the predicted RMS from linear perturbation theory by a factor of 3. While this is an ad-hoc solution, it yields the correct relative velocity distributions on all scales (c.f~Figs.~\ref{fig: 4} and \ref{fig: 14}).


\section{Beyond linear perturbation theory}\label{sec: Beyond linear perturbation theory}
Throughout this work, we have been working within the formalism of linear perturbation theory in order to make the analytical calculations in {\tt SP$\alpha$RTA}. In this appendix, we go beyond linear perturbation theory, and explore the validity of our approximations. Specifically, we test if the statistics of the relative velocity field change significantly when going beyond linear perturbation theory.

For that purpose, we employ a second order Lagrangian perturbation theory (2LPT) simulation from {\tt 21cmFAST}. In 2LPT, parcels are shifted from their initial Lagrangian coordinates to their final Eulerian coordinates according to the displacement vector field, which is expanded to second order corrections~\cite{Bouchet:1994xp, Scoccimarro:1997gr}. The collapse of parcels (i.e. matter) under the influence of gravitational forces then yields the non-linear density field, from which the non-linear peculiar velocity field is derived via the continuity equation. We make our 2LPT simulation on a box of size $150\,\mathrm{Mpc}$ with a cell size of $L_\mathrm{cell}=0.75\,\mathrm{Mpc}$. The final redshift of the simulation, which is considered as the absorption redshift, is $z_\mathrm{abs}=10$.

To gather enough statistics for the numerical relative velocity distributions, we post-process the three output velocity boxes $\mathbf v$ of the 2LPT simulation (one box for each velocity component) as follows. For each $x_\mathrm{em}$ value, we select $n_\mathrm{abs}=5000$ absorption points and compute the straight line-distance between the emitter and absorber via $R_\mathrm{SL}=R_*\left(z_\mathrm{abs}\right)x_\mathrm{em}$. Given this $R_\mathrm{SL}$ value, for each  
absorption point $\mathbf r_\mathrm{abs}$, we find $n_\mathrm{em}=1000$ potential emission points that their separation $r=\left|\mathbf r_\mathrm{em}-\mathbf r_\mathrm{abs}\right|$ from the absorption point satisfies $\left|R_\mathrm{SL}-r\right|\leq1.5L_\mathrm{cell}$. This gives us $5\times10^6$ independent samples of the peculiar velocity vector at absorption and emission points, $\mathbf v_\mathrm{abs}$ and $\mathbf v_\mathrm{em}$, and their relative velocity vector $\mathbf v_\mathrm{rel}=\mathbf v_\mathrm{em}-\mathbf v_\mathrm{abs}$. The parallel component of the velocity vector is found via projection along the separation vector, $v^{||}_\mathrm{rel}=\mathbf v_\mathrm{rel}\cdot\left(\mathbf r_\mathrm{em}-\mathbf r_\mathrm{abs}\right)/r$. Then, for each $x_\mathrm{em}$ value we construct a histogram that collects all of our sampled $\pm v^{||}_\mathrm{rel}$ values\footnote{We take both positive and negative values of the sampled $v^{||}_\mathrm{rel}$ because of the expected symmetry of the distributions around zero, and also to avoid biasing the distributions due to partial double counting of absorption and emission pairs.}. We note that in order to compare apples to apples, it is crucial to divide the velocity output of {\tt 21cmFAST} by $1+z_\mathrm{abs}$, since that output represents the \emph{comoving} velocity field, while in {\tt SP$\alpha$RTA} we work with \emph{proper} velocities.

\begin{figure}
\begin{centering}
\includegraphics[width=\columnwidth]{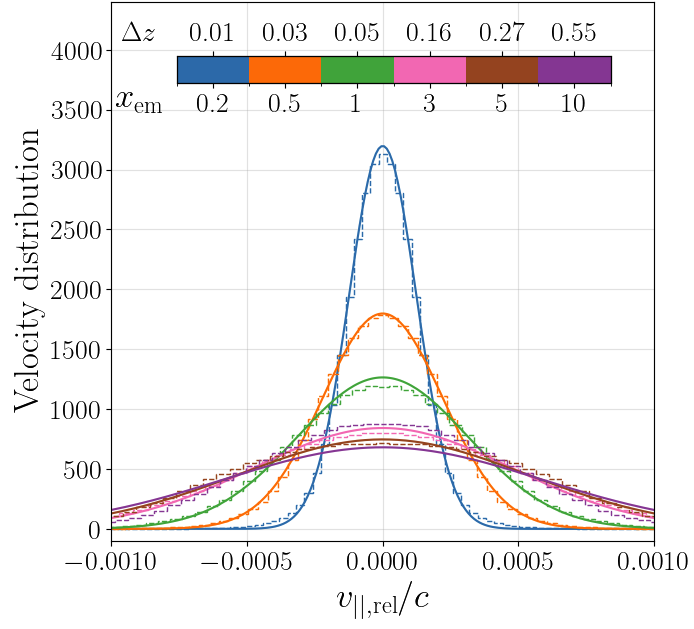}
\caption{Comparison of the relative velocity distribution between linear perturbation theory and 2LPT, at $z_\mathrm{abs}=10$, for different separation scales, as indicated by the $x_\mathrm{em}$ value. \emph{solid} lines correspond to linear perturbation theory, while dashed histograms were obtained from {\tt 21cmFAST}, as detailed in Appendix \ref{sec: Beyond linear perturbation theory}.} 
\label{fig: 14}
\end{centering}
\end{figure}

We show in Fig.~\ref{fig: 14} the 2LPT histograms together with the predicted relative velocity distributions from linear perturbation theory. There is a striking good match between the two for $0.2\leq x_\mathrm{em}<10$, which corresponds to scales of $2\lesssim R_\mathrm{SL}\lesssim100\,\mathrm{Mpc}$ at $z_\mathrm{abs}=10$. On higher scales, e.g. $x_\mathrm{em}=10$, differences can be seen, but this is purely due to lack of sufficient number of independent samples in a periodic box (we have checked that the same differences on large scales occur when running {\tt 21cmFAST} in linear perturbations configuration). At higher redshits, the agreement is expected to be better, since the peculiar velocity is smaller at earlier times and there is a better match with linear perturbation theory. 

We thus conclude that our qualitative and quantitative conclusions in this paper would have not been significantly changed if we had simulated the photon trajectories on a grid with non-linear peculiar velocity field, at least for scales above $2\,\mathrm{Mpc}$, which is comparable to our {\tt 21cmFAST} cell size in Sec.~\ref{sec: The effect of Lya multiple scattering on the 21cm signal}. A more through investigation of the effect of non-linearities on smaller scales is beyond the scope of this paper and we defer it to future work. 

\section{Implementation of $M_\mathrm{MS}$ in {\tt 21cmFAST}}\label{sec: Implementation of M-MS in 21cmFAST}

The MS window function in {\tt 21cmFAST}, $M_\mathrm{MS}\left(kR\right)$, is modeled via Eq.~\eqref{eq: 32}. According to this equation, $M_\mathrm{MS}\left(kR\right)$ can be expressed as an analytic function, $_2F_3$, a hypergeometric function with 2+3 parameters. While the popular {\tt gsl} library \citep{galassi2018scientific} supports the evaluation of several basic hypergeometric functions, it does not support $_2F_3$. The purpose of this appendix is thus to describe how $_2F_3$ was implemented in {\tt 21cmFAST}. To reduce clutter, we define below $x\equiv kR$.

\subsection{Small $x$ limit}\label{subsec: Small x limit}

While all hypergeometric functions, including $_2F_3$, can be represented as a Taylor expansion, the coefficients of $M_\mathrm{MS}\left(x\right)$ can be simplified by considering the infinitely thin shell window function, Eq.~\eqref{eq: 31}.
\begin{flalign}\label{eq: D1}
&\nonumber W_\mathrm{MS}\left(x\right)=&
\\&{_2F_{3}}\left(\frac{2+\alpha}{2},\frac{3+\alpha}{2};\frac{3}{2},\frac{2+\alpha+\beta}{2},\frac{3+\alpha+\beta}{2};-\frac{1}{4}x^2\right).&
\end{flalign}
The special structure of the parameters of this hypergeometric function allows us to use the following identity, 
\begin{flalign}\label{eq: D2}
&\nonumber{_2F_{3}}\left(a,a+\frac{1}{2};\frac{3}{2},d,d+\frac{1}{2};z\right)=\frac{2a-1}{4\left(2d-1\right)\sqrt{z}}\times&
\\&\left[{_1F_{1}}\left(2a-1;2d-1;2\sqrt{z}\right)-{_1F_{1}}\left(2a-1;2d-1;-2\sqrt{z}\right)\right],&
\end{flalign}
where $_1F_1$ is the confluent hypergeometric function. Thus, we may write
\begin{flalign}\label{eq: D3}
&\nonumber W_\mathrm{MS}\left(x\right)=\frac{\alpha+\beta+1}{\alpha+1}\hspace{0mm}\times&
\\&\frac{_{1}F_{1}\left(\alpha+1;\alpha+\beta+1,ix\right)-{}_{1}F_{1}\left(\alpha+1;\alpha+\beta+1,-ix\right)}{2ix}.&
\end{flalign}
This expression for $W_\mathrm{MS}\left(x\right)$ can be seen as a generalization to $W_\mathrm{SL}\left(x\right)=\sin x/x=\left(\mathrm{e}^{ix}-\mathrm{e}^{-ix}\right)/2ix$, namely $\left[1+\beta/\left(\alpha+1\right)\right]{_{1}F_{1}}\left(\alpha+1;\alpha+\beta+1,z\right)$ can be viewed as the MS generalization to $\mathrm{e}^z$. 

We can now use the definition of $_1F_1$,
\begin{equation}\label{eq: D4}
_{1}F_{1}\left(a;b;z\right)\equiv\sum_{n=0}^{\infty}\frac{\left(a\right)_{n}}{\left(b\right)_{n}}\frac{z^{n}}{n!}=\frac{\Gamma\left(b\right)}{\Gamma\left(a\right)}\sum_{n=0}^{\infty}\frac{\Gamma\left(a+n\right)}{\Gamma\left(b+n\right)}\frac{z^{n}}{n!},
\end{equation}
where $\left(a\right)_{n}$ is the Pochhammer symbol, which can be expressed with gamma functions. By combining Eq.~\eqref{eq: D3} and Eq.~\eqref{eq: D4}, it is then straightforward to show that
\begin{flalign}\label{eq: D5}
&\nonumber W_\mathrm{MS}\left(x\right)=\sum_{n=0}^{\infty}\frac{\Gamma\left(\alpha+\beta+2\right)}{\Gamma\left(\alpha+\beta+2n+2\right)}&
\\&\hspace{20mm}\times\frac{\Gamma\left(\alpha+2n+2\right)}{\Gamma\left(\alpha+2\right)}\frac{\left(-1\right)^{n}x^{2n}}{\left(2n+1\right)!}.&
\end{flalign}
The desired finite-shell window function can then be achieved using Eq.~\eqref{eq: 12}. As integrating the RHS of Eq.~\eqref{eq: D5} is trivial, we find that 
\begin{flalign}\label{eq: D6}
&\nonumber M_\mathrm{MS}\left(x\right)=\sum_{n=0}^{\infty}\frac{\Gamma\left(\alpha+\beta+2\right)}{\Gamma\left(\alpha+\beta+2n+2\right)}&
\\&\hspace{20mm}\times\frac{\Gamma\left(\alpha+2n+2\right)}{\Gamma\left(\alpha+2\right)}\frac{3}{2n+3}\frac{\left(-1\right)^{n}x^{2n}}{\left(2n+1\right)!}.&
\end{flalign}

For numerical purposes, in order to avoid the evaluation of the gamma functions, it is best to express $M_\mathrm{MS}\left(x\right)=\sum_{n=0}^{\infty}A_{n}$, with $A_0=1$ and the following recursion relation,
\begin{equation}\label{eq: D7}
\frac{A_n}{A_{n-1}}=-\frac{1}{\left(1+\frac{\beta}{\alpha+1+2n}\right)\left(1+\frac{\beta}{\alpha+2n}\right)}\frac{x^{2}}{2n\left(2n+3\right)}.
\end{equation}

\subsection{Large $x$ limit}\label{subsec: Large x limit}

While Eq.~\eqref{eq: D6} is exact for all $x>0$, it is numerically unstable when $x\gg 1$. In this limit, we can use the asymptotic approximation to $_2F_3$, which in our case reads
\begin{flalign}\label{eq: D8}
&\nonumber M_\mathrm{MS}\left(x\right)\approx\frac{3}{4}\frac{\Gamma\left(\frac{\alpha+\beta+2}{2}\right)\Gamma\left(\frac{\alpha+\beta+3}{2}\right)}{\Gamma\left(\frac{\alpha+2}{2}\right)\Gamma\left(\frac{\alpha+3}{2}\right)}&
\\&\nonumber\times\Bigg\{\frac{\pi\Gamma\left(\frac{\alpha+2}{2}\right)\left(\frac{x}{2}\right)^{-\alpha-2}}{\Gamma\left(\frac{3-\alpha}{2}\right)\Gamma\left(\frac{\beta}{2}\right)\Gamma\left(\frac{\beta+1}{2}\right)}-\frac{2\pi\Gamma\left(\frac{\alpha+3}{2}\right)\left(\frac{x}{2}\right)^{-\alpha-3}}{\Gamma\left(\frac{2-\alpha}{2}\right)\Gamma\left(\frac{\beta-1}{2}\right)\Gamma\left(\frac{\beta}{2}\right)}&
\\&\nonumber+\left[\left(1+\left(\alpha-1\right)\beta\right)\frac{\sin\left(x-\frac{\pi\beta}{2}\right)}{x}-\cos\left(x-\frac{\pi\beta}{2}\right)\right]&
\\&\hspace{65mm}\times\left(\frac{x}{2}\right)^{-\beta-2}\Bigg\}.&
\end{flalign}

In {\tt 21cmFAST}, we use $x=30$ as the threshold value to determine if we use Eq.~\eqref{eq: D6} or Eq.~\eqref{eq: D8}. However, if $\alpha/\beta\gg1$, the asymptotic approximation is not adequate at $30\lesssim x\lesssim100$ and the asymptotic formula diverges at this range. Yet, $\alpha\gg\beta$ corresponds to the SL limit. Since, as a rule of thumb, $\left|M_\mathrm{MS}\left(x\right)\right|<\left|M_\mathrm{SL}\left(x\right)\right|$ at high $x$ values (c.f.~\ref{fig: 8}), we thus apply the following logic; if $\left|M_\mathrm{MS}\left(x\right)\right|>\left|M_\mathrm{SL}\left(x\right)\right|$, when $M_\mathrm{MS}\left(x\right)$ is evaluated with the asymptotic approximation of Eq.~\eqref{eq: D8}, then we set $M_\mathrm{MS}\left(x\right)=M_\mathrm{SL}\left(x\right)$. We have compared this scheme to evaluate $_2F_3$ with a more precise evaluation by the {\tt mpmath} package \citep{mpmath} and found an excellent agreement between the two methods.

\bibliography{paper.bib}

\end{document}